\documentclass[useAMS,usenatbib]{mn2e}
\usepackage{graphicx}
\usepackage{latexsym}
\usepackage{xspace}

\title[Simultaneous Formation of Massive Stars \& Clusters]{The Simultaneous Formation of Massive Stars and Stellar Clusters}

\author[Smith, Longmore \& Bonnell]{Rowan J. Smith$^{1}$ \thanks{Email: rjs22@st-andrews.ac.uk}, Steven Longmore$^{2}$, Ian Bonnell$^{1}$\\
$^1$ SUPA, School of Physics \& Astronomy, University of St Andrews, North Haugh, St Andrews, Fife, KY16 9SS, UK \\
$^2$ Harvard-Smithsonian Center for Astrophysics, 60 Garden Street, Cambridge, MA 02138, USA
 }

\begin{document}

\pagerange{\pageref{firstpage}--\pageref{lastpage}} \pubyear{2008}

\maketitle

\label{firstpage}

\def\solmas{{M$_\odot$}}
\def\solm{{M_\odot}}
\def\mnras{MNRAS}
\def\apj{ApJ}
\def\aap{A\&A}
\def\apjl{ApJL}
\def\apjs{ApJS}
\def\bain{BAIN}
\def\pasp{PASP}
\def\araa{ARA\&A}
\def\ga{\sim}
\def\nat{Nature}
\def\aj{AJ}

%some simplifying definitions

\newcommand{\eq}{Equation }
\newcommand{\fig}{Figure }
\newcommand{\msun}{\,M$_{\odot}$ }
\newcommand{\gcmc}{\,g\,cm$^{-3}$}
\newcommand{\kms}{\,kms$^{-1}$}
\newcommand{\tab}{Table }
\newcommand{\gcms}{\,g\,cm$^{-2}$\xspace}

\begin{abstract}
We show that massive stars and stellar clusters are formed simultaneously, the global evolution of the forming cluster is what allows the central stars to become massive. We predict that massive star forming clumps, such as those observed in \citet{Motte07}, contract and grow in mass leading to the formation of massive stars. This occurs as mass is continually channeled from large radii onto the central proto-stars, which can become massive through accretion. Using SPH simulations of massive star forming clumps in a Giant Molecular Cloud, we show that clumps are initially diffuse and filamentary, and become more concentrated as they collapse. Simulated interferometry observations of our data provide an explanation as to why young massive star forming regions show more substructure than older ones. The most massive stars in our model are found within the most bound cluster. Most of the mass accreted by the massive stars was originally distributed throughout the clump at low densities, and was later funneled to the star due to global in-fall. Even with radiative feedback no massive pre-stellar cores are formed. The original cores are of intermediate mass and gain their additional mass in the proto-stellar stage. We also find that cores which form low mass stars exist within the volume from which the high mass stars accrete, but are largely unaffected by this process.
\end{abstract}
%This needs looked at again at the end


\section{Introduction}
\label{sec:introduction}

Massive stars are almost universally formed in star clusters \citep{Lada03} and so the physical processes involved in forming clusters must be intrinsically linked to high mass star formation. Recent observations of infra-red dark clouds (IRDC's) \citep{Egan98,Carey98,Carey00,Simon06,Rathborne06} and high-mass proto-stellar objects (HMPO's) \citep{Beuther02,Sridharan02,WilliamsSJ04,Faundez04} have begun to probe the earliest stages of the massive clumps from which star clusters are formed.

So how does high mass star formation proceed within the complicated environment of of a forming cluster? Is the mass that forms the massive star gathered before, during or after the cluster formation? This is particularly relevant as regards the evolution and interactions of the pre-stellar gas cores thought to be the precursors of star formation.  

Due to the hierarchical structure of molecular clouds, boundaries are often arbitrary \citep[e.g.][]{Smith08} leading to confusion in the
literature over terms which define discrete entities such as clumps and cores. \citet{Williams00} adopt the terminology that a `clump'
typically contains $50-500$ M$_{\odot}$ within $0.3-3$ pc, and a `core' contains $0.5-5$ M$_{\odot}$ within $0.03-0.2$ pc. A summary of typical properties of clumps and cores is shown in Table \ref{tab:defn}. Effectively we use the terms to simply denote
different scales of structure. Clumps are regions of enhanced density within a molecular cloud, which will typically form stellar
clusters. Cores are density condensations smaller than a clump which have a gravitational potential distinct from their environment and do not contain any smaller scale structure which is already bound.

\begin{table}
\centering
\caption{Typical properties of clouds, clumps and cores (adapted from \citealt{Bergin07}.)}
\begin{tabular}{l c c c }
    \hline
    \hline
     & Cloud & Clump & Core \\
    \hline
    Size (pc) & $2-15$ & $0.3-3$ & $0.03-0.2$\\
    Mass ($M_{\odot}$) & $10^{3}-10^{4}$ & $50-500$ & $0.5-5$\\
    Mean density (cm$^{-3}$) & $50-500$& $10^{3}-10^{4}$ &$10^{4}-10^{5}$\\			
    Velocity Extent ($km s^{-1}$) & $2-5$ & $0.3-3$ & $0.1-0.3$\\
    Gas Temperature (K) & $\sim 10$ & $10-20$& $8-12$\\
    \hline
\end{tabular}
\label{tab:defn}
\end{table}

\citet{Motte07} carried out an unbiased survey of Cygnus X to identify the earliest high mass star forming complexes. They found evidence of star formation in all their embedded cores, and were unable to find a massive pre-stellar clump, leading to the conclusion that these objects were either extremely short lived ($<10^{3}$ years), or did not exist. Similarly, observations of massive cores ( e.g. \citealp{Rathborne05,Pillai06,Andre08}) have always found signatures that star formation was already underway. \citet{Marseille08} find a possible pre-stellar massive core, but any mid-IR emission from it is confused by a nearby source. In conclusion, true pre-stellar massive cores remain elusive.

Due to the lack of pre-stellar massive cores, \citet{Motte07} proposed that the precursors of Class 0 massive proto-stars must be the larger starless clumps which would form them by collapsing supersonically. Dynamic collapse of HMPO's to form stars was also proposed by \citet{Beuther02} to explain their observed line-widths. Further, \citet{Peretto06,Peretto07} found that the massive cluster forming clump NGC 264-C was collapsing along its axis in accordance with its dynamical timescale, and therefore channelling mass towards the Class 0 object at its centre.

This dynamical collapse of clumps during the star-formation process is in agreement with the proposal of \citet{Elmegreen00} that star formation basically takes place in a crossing time and that cloud lifetimes are short. \citet{Tan06} present a contrasting view where clusters form quasistatically, although this is partially based on a large estimate for the age of the Orion nebula cluster. \citet{Hillenbrand97}, however, found that the mean stellar age within the Orion nebula was below $1$ Myr albeit with a few older outliers in the range $1-10$ Myr. \citet{Hartmann03} argue that these outliers are accounted for by a combination of uncertainties in the stellar birth line where stars appear on the H-R diagram and foreground contamination.

Additionally, observations of a velocity gradient within the Orion nebula by \citet{Furesz08,Tobin09} suggest that the cluster may currently be in a state of sub-virial collapse, and due to the kinematic correlation between stellar and gaseous components must be young. The focussing power of gravity to produce structures has been further highlighted by \citet{Hartmann07}, who suggest that the Orion Nebula Cluster itself could have been produced from the large scale collapse of the Orion A cloud.

%In observations of hot molecular cores \citet{Longmore07} were able to separate their sample into young, intermediate and old objects, using temperature and the presence of methanol masers. When the cores were followed up using the the SMA interferometer, the young objects were found to contain considerable substructure in comparison with the older cores (Longmore 2009, in preparation). \citet{Zhang09} similarly found that the younger of two massive cores surveyed with the SMA contained the most substructure.

There are two main theories of massive star formation; the first of which is basically a scaled up version of low mass star formation, where massive stars form from well defined massive cores supported by turbulence \citep{McKee03}. The difficulty with this model is that it presupposes the existence of massive prestellar cores that have somehow evaded fragmentation during their formation stages \citep{Dobbs05}. \citet{Krumholz06} suggest that radiative feedback can limit the fragmentations but as we shall see here, radiative feedback does not result in the formation of massive prestellar cores.

Alternatively, in the competitive accretion scenario \citep{Zinnecker82,Bonnell01,Bonnell04} cores are the `seeds' of star formation, the most massive of which have a larger gravitational radius, and are thus more successful at accreting additional mass, and so grow into massive stars. There are a few common misapprehensions about this theory. Firstly, the protostars which become massive do not generally have high velocities with respect to the cloud they inhabit. They tend to stay at the centre of gravitational potential of the forming star cluster which they help define. Secondly, they accrete material via two mechanisms. There is a contribution from Bondi-Hoyle accretion, but when the velocity relative to the system is low the accretion is mainly regulated by the tidal field. Thirdly, there is no requirement for stellar mergers \citep{Bonnell98}.

At later evolutionary stages, \citet{Keto07} has shown that accretion can continue to form massive stars, even when they have begun to ionise their surroundings. In this model a hypercompact HII region is formed around the massive stars as they grow by accretion. As the star grows in mass, outflows form and their opening angle increases, however even when the star is extremely massive, accretion still proceeds around an equatorial disk.

In this paper we shall show how massive stars are formed within a dynamic clump which is forming a stellar cluster, with particular attention to the cores within it. In Section \ref{sec:sim} we outline our numerical simulation and describe a simple approximation for radiative heating. In Section \ref{sec:tevo} we show how dynamic collapse causes the star forming clump to evolve from a diffuse filamentary structure to a more massive concentrated structure which is brighter in dust continuum emission. We also present simulated interferometry images of our data and compare these to observations. We discuss our results in the context of global collapse and accretion in Section \ref{sec:interaction}, where we show that the clump potential channels mass onto its centre, where the proto-stars with the greatest gravitational radius accrete it. This means that the global evolution of the forming star cluster as a whole has a direct link to the massive stars it forms. We also show that low mass cores close to the central massive star are unaffected by this process. Finally in Section \ref{sec:conc} we give our conclusions.

\section{The Simulation}
\label{sec:sim}
We use the smoothed particle hydrodynamics (SPH) method to follow the evolution of a $10^{4}$ M$_{\odot}$ cloud over
$1.02$ free-fall times or $\approx 6.6 \times 10^{5}$ years. The initial conditions are the same as reported in \citet{Bonnell08}. The cloud is cylindrical in form, with a length of $10$ pc and a radius of $3$ pc. The cloud contains a local density gradient causing the ends of the cylinder to have initial gas densities that are $33 \%$ higher/lower than the average density. The gas has internal turbulence following a Larson-type $P(k) \sim k^{-4}$ power law and is normalised so that the total kinetic energy balances the total gravitational energy in the cloud. The density gradient then results in one end of the cloud being over bound (still super virial) while the other end of the cloud is unbound.

The cloud is made up of $15.5$ million SPH particles on two levels to maximise the numerical resolution in regions of interest. The regions of interest are determined from an initial low resolution run \citep{Smith09} of 5 million SPH particles with mass resolution of $0.15$ \solmas. Regions requiring high resolution were identified from the SPH particles that underwent star formation and formed sink particles (see below) or were subsequently accreted by these sink particles. These particles were replaced in the initial conditions with 9 lower-mass particles, conserving the mass and kinetic energy of the initial conditions, but now with a mass resolution of $0.0167$\solmas. The simulation was re-run from the initial conditions with this higher resolution.

The simulation treats the thermal content of the cloud through a barotropic equation of state designed to mimic a cooling evolution when line cooling dominates at low densities followed by an isothermal evolution at higher densities when dust cooling dominates \citep[eg.][]{Larson05}. The equation of state is given by \begin{equation}
P = k \rho^{\gamma}
\end{equation}
where
\begin{equation}
\begin{array}{rlrl}
\gamma  &=  0.75  ; & \hfill &\rho \le \rho_1 \\
\gamma  &=  1.0  ; & \rho_1 \le & \rho  \le \rho_2 \\
\gamma  &=  1.4  ; & \hfill \rho_2 \le &\rho \le \rho_3 \\
\gamma  &=  1.0  ; & \hfill &\rho \ge \rho_3, \\
\end{array}
\end{equation}
and $\rho_1= 5.5 \times 10^{-19} {\rm g\ cm}^{-3} , \rho_2=5.5 \times 10^{-15} {\rm g\ cm}^{-3} , \rho_3=2 \times 10^{-13} {\rm g\ cm}^{-3}$. At densities above $\rho_3$, sink particles \citep{Bate95} are used to model star formation provided the region is bound and collapsing.

Additionally, we approximate the radiative feedback from the newly formed stars by way of a grid of previously computed Monte Carlo radiative transfer models of young stars \citep{Robitaille06}. We derive a one-dimensional temperature profile from the youngest of these models as a function of stellar mass and distance from the star. This is a very rough estimate of the radiative feedback which, if anything, should overestimate the gas temperatures. Nevertheless, it gives us an estimate of the emission expected in regions of massive star formation. From these models we set the temperature due to the radiative feedback as
\begin{equation}
\begin{array}{rlrl}
T(r) & =  100. \left(\frac{m}{ 10_{\odot}}\right)^{0.35} \left(\frac{r}{1000 AU}\right)^{-0.45}   K ; &  &m < 10 M_{\odot}, \\
T(r) & =  100. \left(\frac{m}{ 10_{\odot}}\right)^{1.11} \left(\frac{r}{1000 AU}\right)^{-0.5}   K  ; &  &m > 10 M_{\odot}\\
\end{array}
\end{equation}
The gas temperature around the young stars is set to be the maximum of the temperature from either the the dust/line cooling equation of state or the radiative feedback. This ensures a maximal effect from the radiation.

\begin{table}
	\centering
	\caption{The massive clump properties recorded at the beginning ($3.53\times10^{5}$ yrs) and end ($5.9\times10^{5}$ yrs) of the analysed period. The first mass $M$ is that found within a 1 pc radius of the central sink. The second is that within a cylindrical column of radius 1pc centered on the same sink. The mean gas density is denoted by $\bar{\rho_{g}}$, and max. M$_{s}$ and tot. M$_{s}$ represent the maximum sink mass within the clump and the total sink mass respectively.}
	\begin{tabular}{l c c c c c}
	\hline
	\hline
	Clump & M & M$_{2D}$ & $\bar{\rho_{g}}$ & max. M$_{s}$ & tot. M$_{s}$ \\
	 &[\msun]  & [\msun]& [gcm$^{-3}$] &[\msun] & [\msun] \\
	\hline
	\textit{beginning}\\
	Alpha & $893$ & $1528$ & $1.1\times10^{-18}$ &$0.85$ &$3.10$ \\
	Beta & $882$ & $1516$ & $4.0\times10^{-19}$&$1.11$ & $2.24$ \\
	Gamma & $1034$ & $1985$ & $7.6\times10^{-19}$&$0.58$ &$1.84$ \\
	\hline
	\textit{end}\\
	Alpha & $987$ & $1412$ & $8.0\times10^{-18}$ &$29.2$ &$361.4$ \\
	Beta & $995$ & $1882$ & $8.8\times10^{-18}$&$11.3$ & $189.2$ \\
	Gamma & $1127$ & $1993$ & $5.0\times10^{-18}$&$12.6$ &$243.9$ \\
	\hline
	\end{tabular}
	\label{Regprops}
\end{table}

\begin{table}
\centering
\caption{The binding energy of the three clumps at the beginning ($3.53\times10^{5}$ yrs) and end ($5.9\times10^{5}$ yrs) of the analysed period. The average radial velocity of the sph gas particles is given by $\bar{v_{r}}$, where negative values show inward motion. The binding of the clumps is given by E$_{rat}$=E$_{p}$/E$_{kin}$+E$_{therm}$, and the absolute magnitude of the potential energy is shown by E$_{p}$. For E$_{rat2}$, the kinetic energy was calculated without including any inward velocities.}
	\begin{tabular}{l c c c c}
	\hline
	\hline
	Clump &$\bar{v_{r}}$ & E$_{rat}$ & E$_{rat2}$& E$_{p}$\\
	&[kms$^{-1}$] & & & [erg]\\
	\hline
	\textit{beginning}\\
	Alpha & $-0.45$ & $3.4$ & $18.8$ & $1.26\times10^{-47}$\\
	Beta & $-0.62$ & $0.8$ & $4.6$ & $8.64\times10^{-46}$\\
	Gamma & $-0.44$ & $1.8$& $11.1$ & $1.08\times10^{-47}$\\
	\hline
	\textit{end}\\
	Alpha & $-1.62$ & $1.09$ & $3.0$ & $4.78\times10^{-47}$\\
	Beta & $-1.16$ & $0.57$ & $3.0$& $2.49\times10^{-47}$\\
	Gamma & $-0.19$ & $1.61$ & $4.5$ & $2.64\times10^{-47}$\\
	\hline
	\end{tabular}
	\label{Bindprops}
\end{table}

\section{Massive Clump Evolution}
\subsection{Time Evolution}
\label{sec:tevo}
\begin{figure*}
\begin{center}
\begin{tabular}{c c c}
\includegraphics[angle=0,width=2.2in]{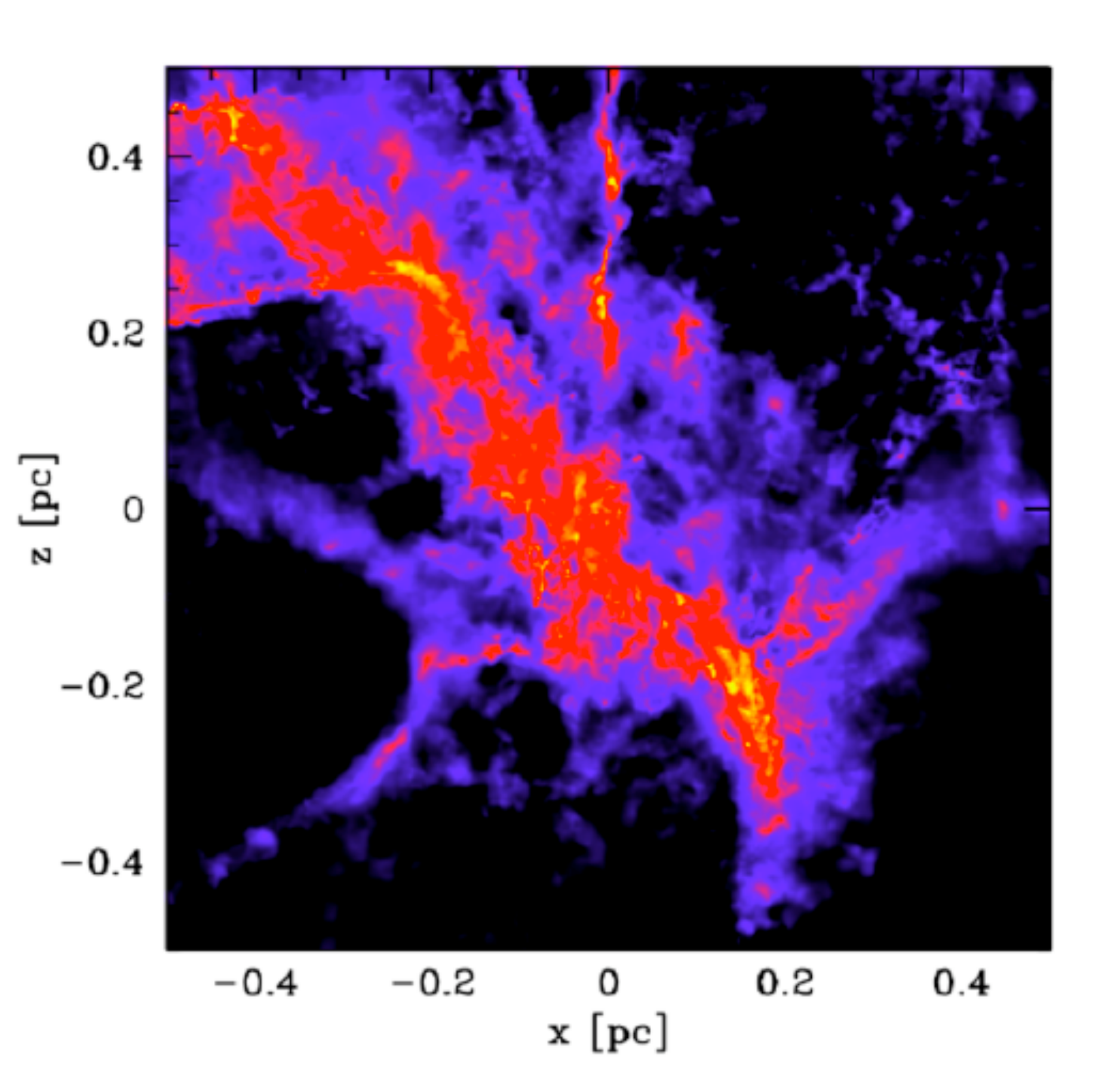}
\includegraphics[angle=0,width=2.2in]{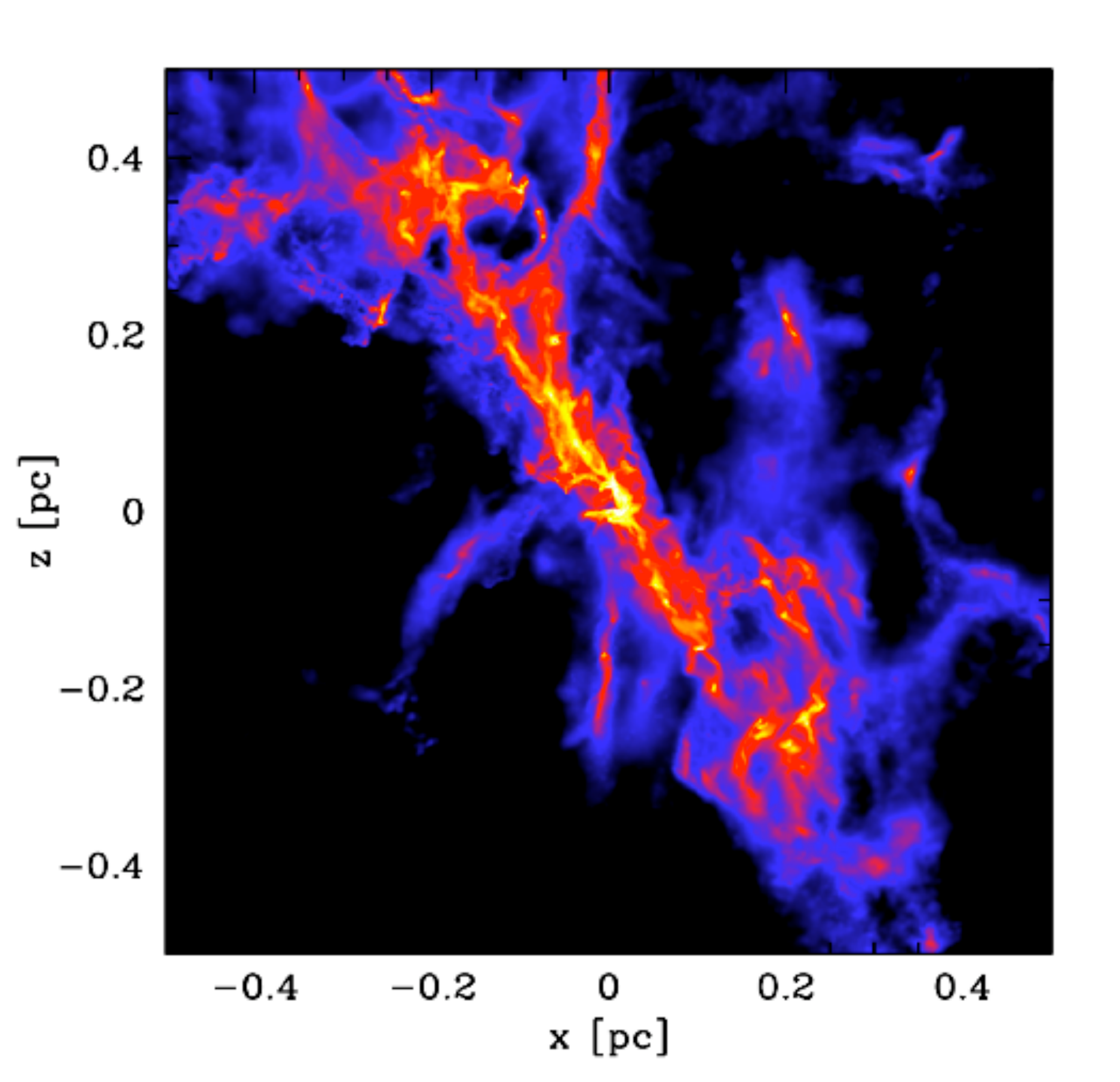}
\includegraphics[angle=0,width=2.2in]{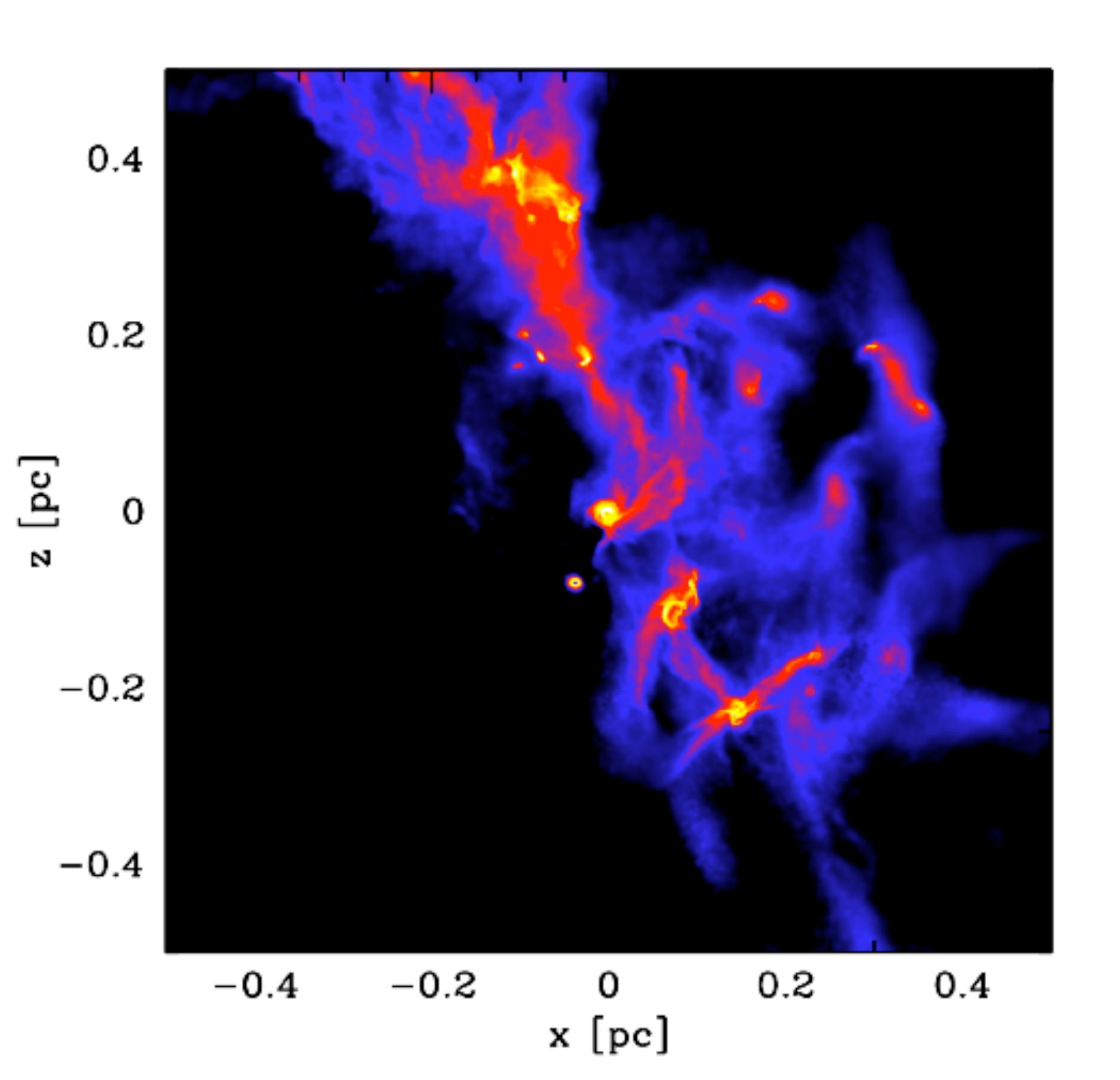}\\
\includegraphics[angle=0,width=2.2in]{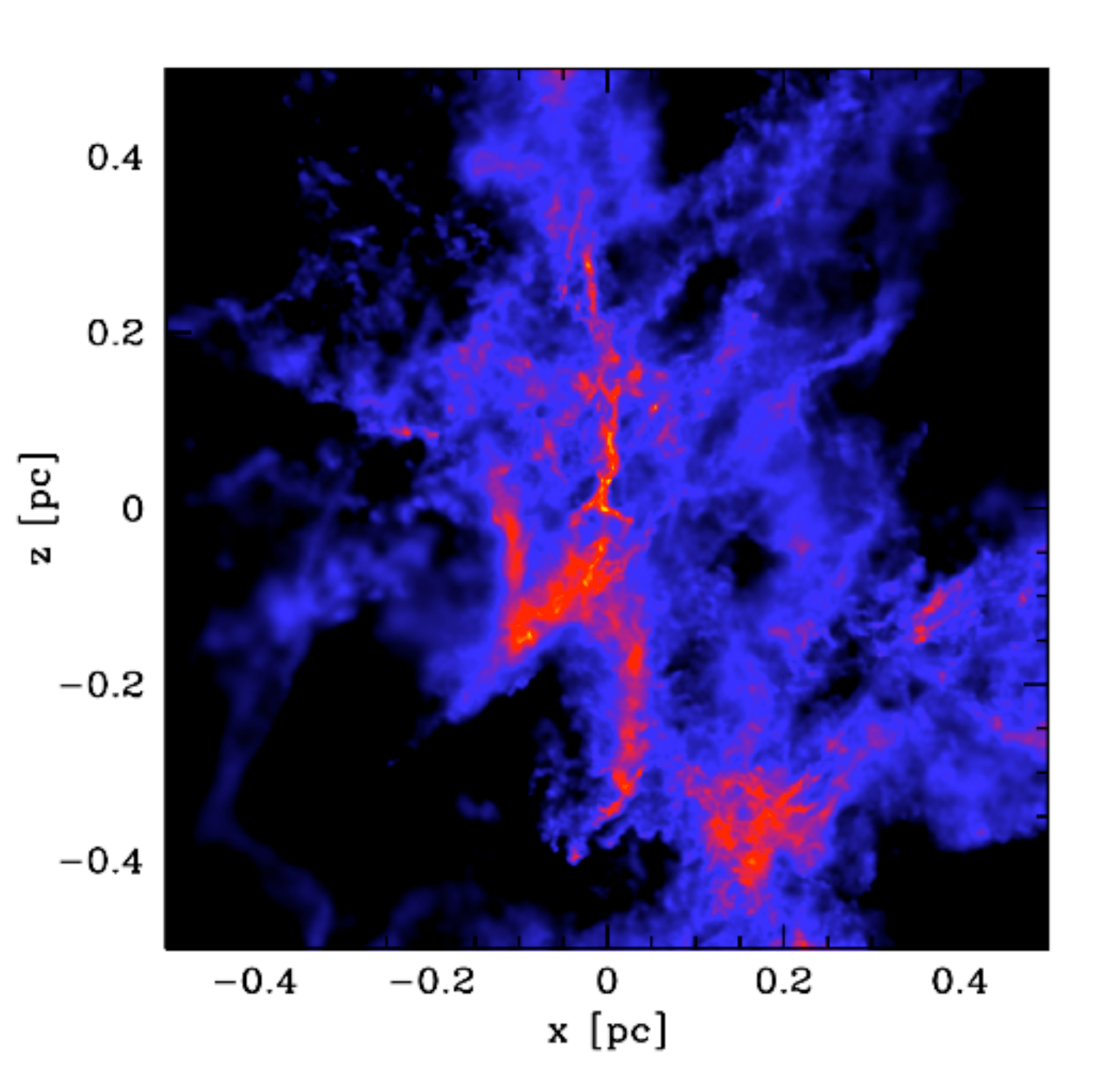}
\includegraphics[angle=0,width=2.2in]{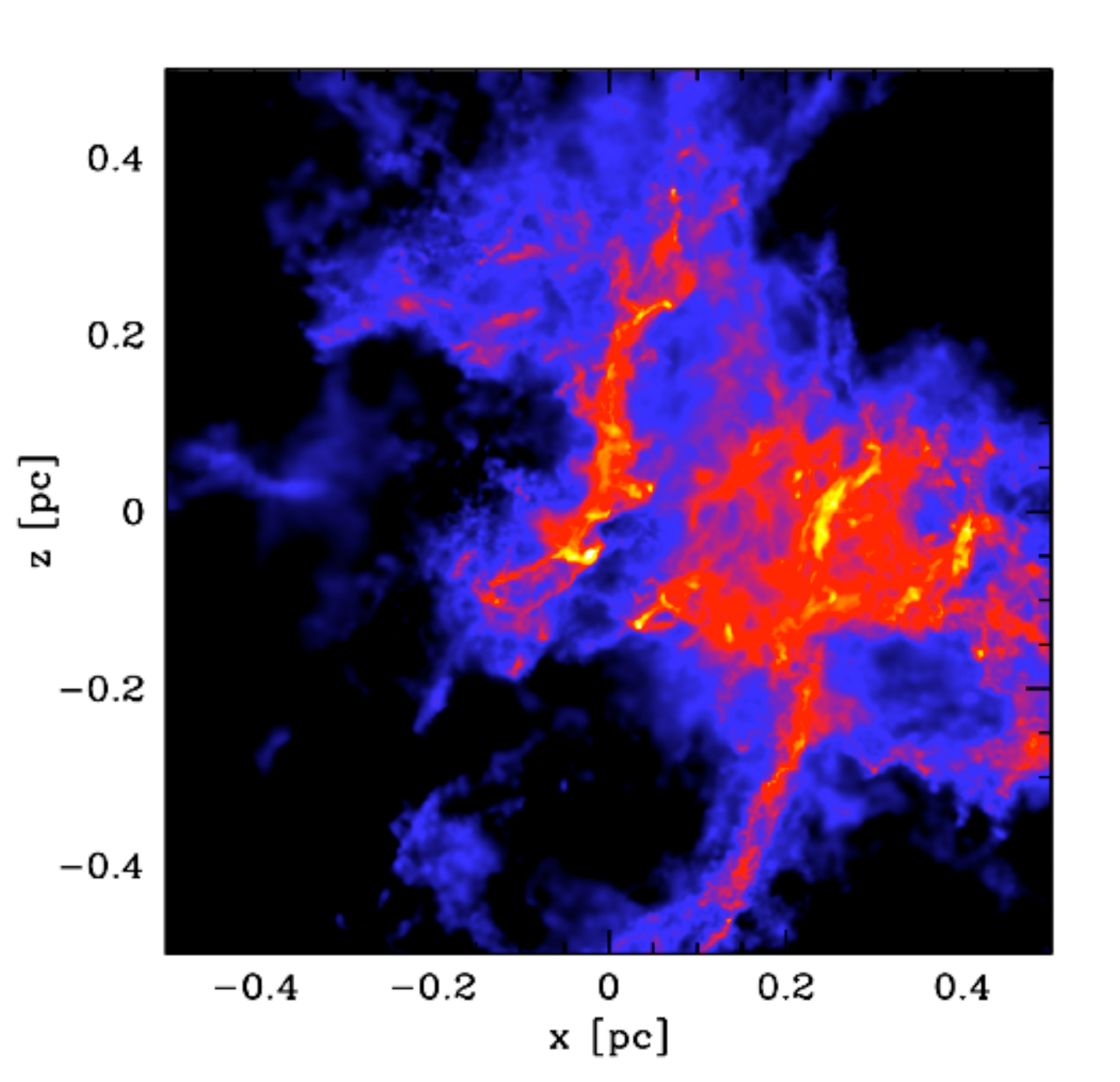}
\includegraphics[angle=0,width=2.2in]{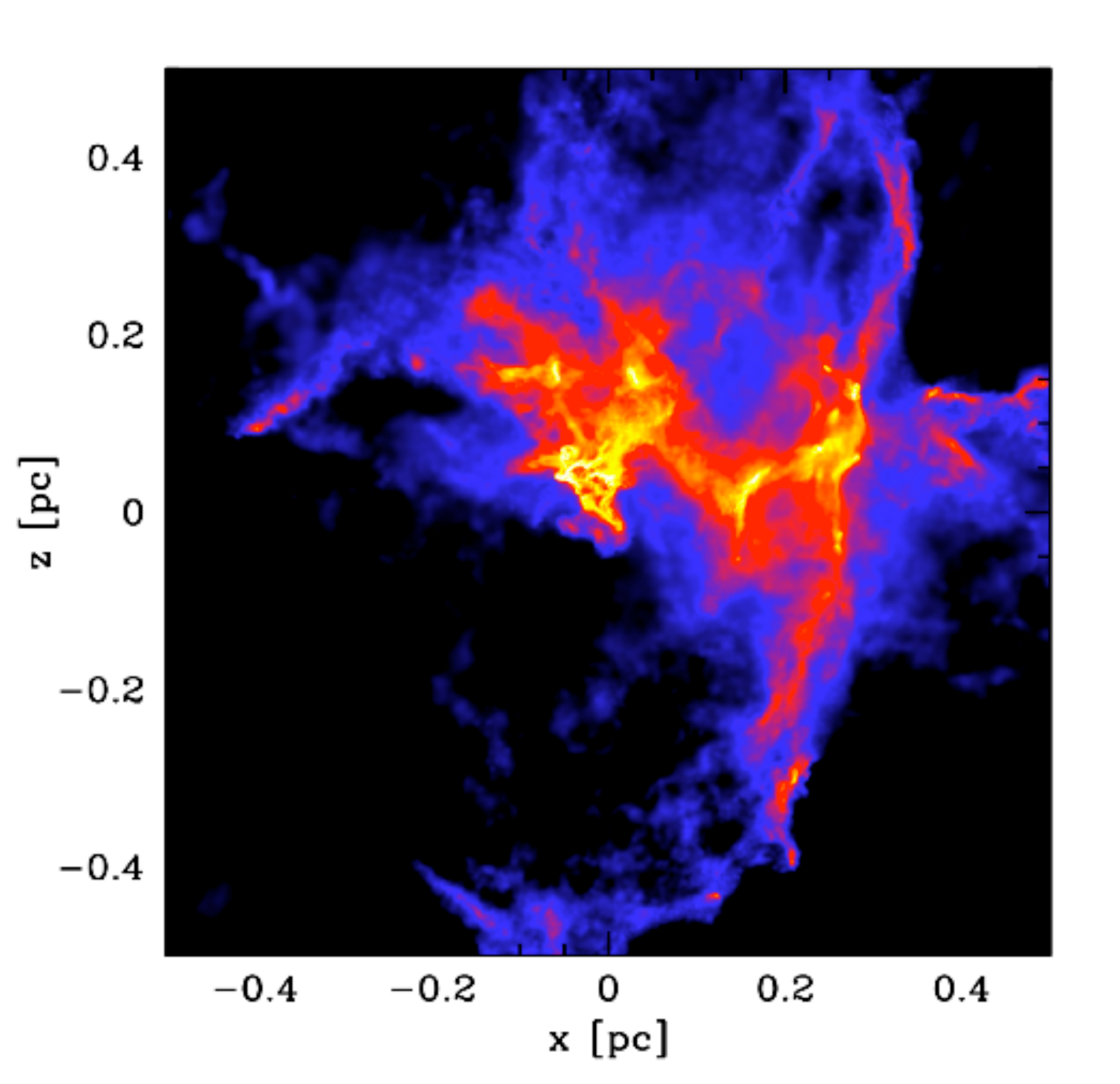}\\
\includegraphics[angle=0,width=2.2in]{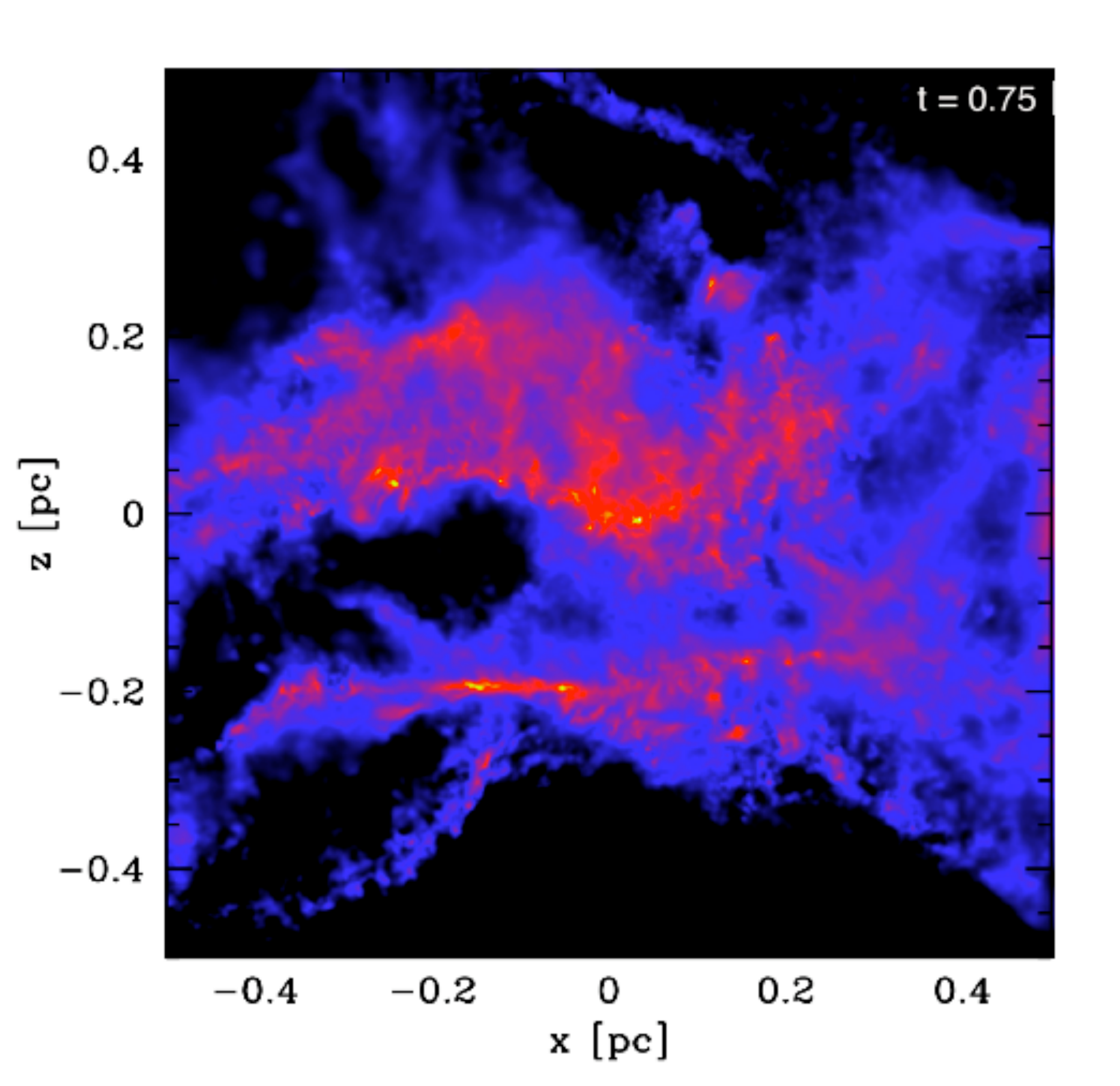}
\includegraphics[angle=0,width=2.2in]{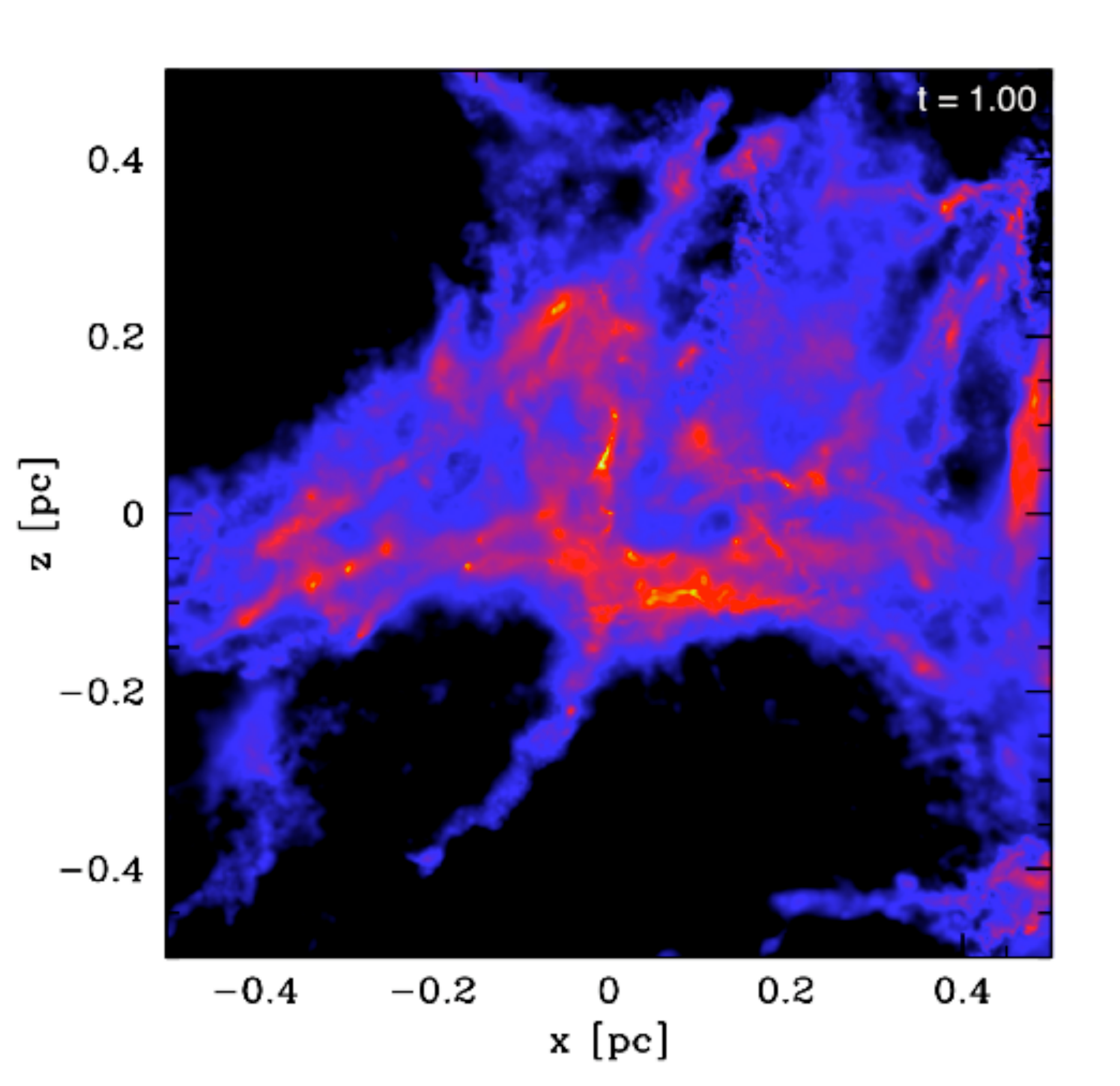}
\includegraphics[angle=0,width=2.2in]{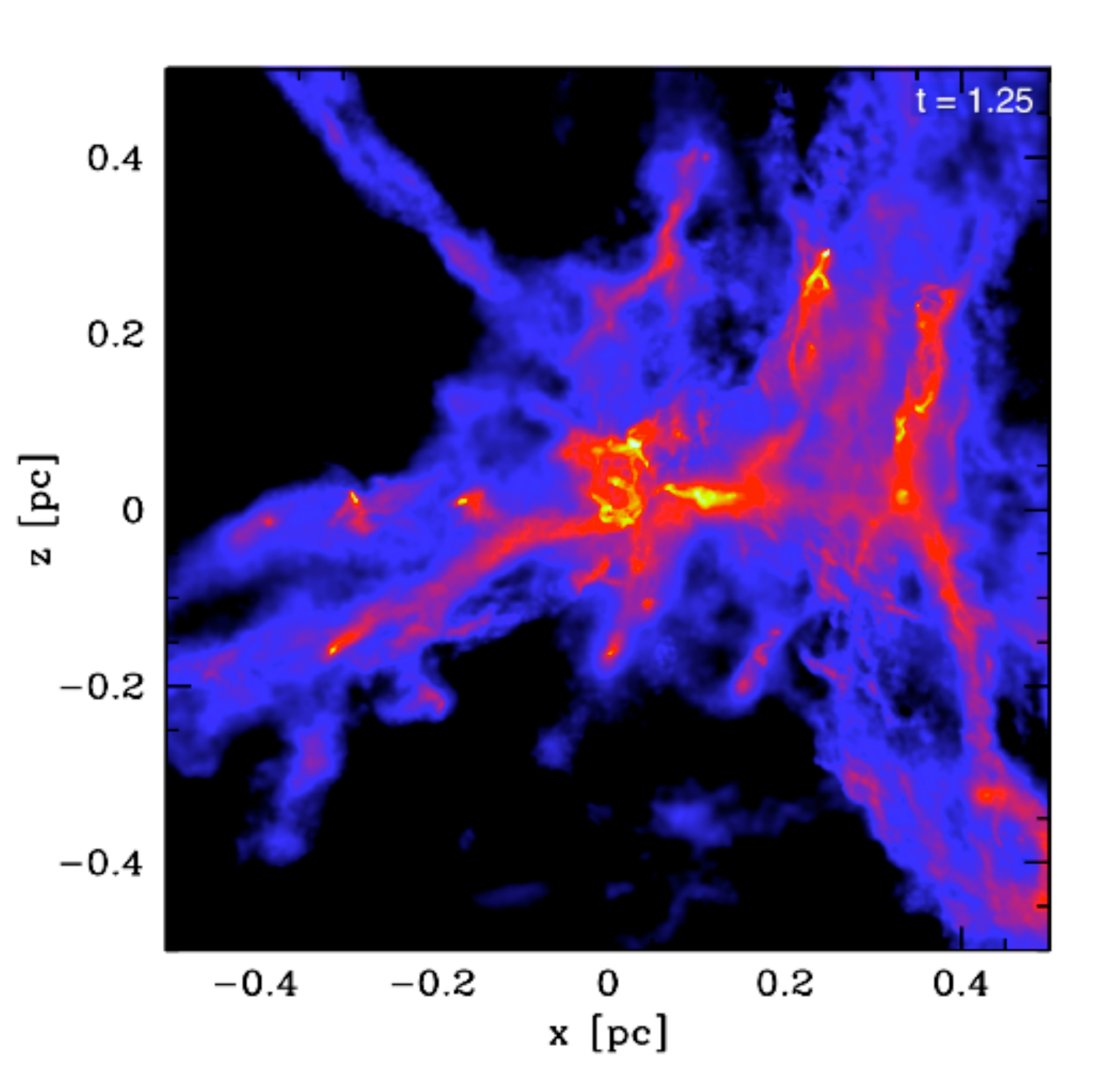}\\
\end{tabular}
\end{center}
\caption{The evolution of the centre of clumps Alpha \textit{top}, Beta \textit{middle} and Gamma \textit{bottom}. The snapshots shown are at \textit{left} $0.75$ t$_{dyn}$ ($3.53\times10^{5}$ yrs),  \textit{middle} $1$ t$_{dyn}$ ($\sim4.7\times10^{5}$ yrs) and  \textit{right} $1.25$ t$_{dyn}$ ($5.9\times10^{5}$ yrs) respectively. The colour scale denotes column densities from $0.05$ \gcms to $5$ \gcms. The structure becomes more compact with time, and in the case of clump Alpha decreases in substructure.}
\label{app}
\end{figure*}

We consider three regions of star formation, each of which resembles a single clump at some point when viewed at a low resolution. We calculate the global properties of these regions by simply including all the material within a radius of $1$ pc from the largest sink. Although this is a simplistic definition, we prefer it to a clumpfinding approach for several reasons. Firstly, the boundaries found from clumpfinding are extremely subjective \citep{Smith08,Kainulainen09} and secondly, defining an absolute spatial scale allows an unambiguous comparison of the physical properties of the studied regions. We name these `clumps', Alpha, Beta and Gamma and Table \ref{Regprops} shows their properties.

The mass in the clumps is high in all cases. This is partly due to our decision to use a large clump radius, so that the analysis is not complicated by additional mass entering the region at later times, but also due to the requirement that to form a large cluster you need a lot of mass. A 2D projected measurement increases the clump masses by $50-100\%$ compared to the 3D case due to contamination from along the line of sight. This highlights the problems of determining masses from only 2D information. In every case the mean gas density in the clumps increases with time due to collapse. The most massive sink is formed in clump Alpha despite the fact that it is not the most massive clump. However this clump does contain the largest total mass in sinks and has the deepest potential well.

Table \ref{Bindprops} outlines the in-fall velocities and binding energies of the clumps at the beginning and end of the simulation. All of the studied regions exhibit significant supersonic in-fall motions, with the exception of clump Gamma at the end of the simulation, which is roughly sonic, ($c\approx0.2$ kms$^{-1}$). This is in agreement with the observation of \citet{Motte07} that rapid supersonic inward motions are required to enhance clump densities to the values seen in massive proto-stellar cores. Typical infall here is a few times the sound speed, which equates to lifetimes of the order of a million years. 

The relative binding of the clumps provides an explanation for the location of the most massive sinks.  At the beginning  of our analysis; clump Alpha is $3.4$ times over-bound, clump Beta appears unbound, and clump Gamma is $1.8$ times over-bound. The binding decreases at the end of the studied period due to the large increase in kinetic energy from randomised gas motions during the collapse process, and heating from the sinks. It is surprising that clump Beta appears unbound, yet is collapsing. This was due to the absolute magnitude of the gas velocities being used to calculate the kinetic energy, despite the fact that in-fall velocities are not supportive. To address this a second energy ratio, $E_{rat2}$, was calculated where the inward velocities were excluded (Vazques-Semadeni, private communication). Using this measure of binding all of the clumps are significantly bound, particularly clump Alpha (which contains the largest sink) which is now $18.8$ times over-bound at the beginning of the studied period.
 
When the magnitude of the potential energy of the clumps is considered directly it is seen that once again Clump Alpha has the greatest potential, followed by Gamma and then Beta. This is the same order as the clumps which contain the most massive sinks, and for the clumps with the greatest total mass in sinks. This suggests that the star formation process is most efficient where the potential energy of the clump as a whole is highest, and this is reflected both in the efficiency of forming a stellar cluster and in the efficiency of forming a massive star. This provides a natural explanation for the link between total stellar mass and the most massive star outlined in \citep{Bonnell04}.

In Figure \ref{app} the column density of gas within the clumps is shown at three snapshots in their evolution. For clarity, sink particles denoting sites of star formation are not shown. The central region of clump Alpha is shown in the top row of Figure \ref{app} at $0.75$, $1$, and $1.25$ dynamical times (t$_{dyn}=4.7\times10^{5}$ yrs). It has a filamentary geometry and the self gravity of the ends of the filament is strong enough for them to collapse independently. However, both objects have inward velocities and are collapsing towards the centre. Notice how the later structure is less filamentary and contains less substructure. Each of the large condensations along the filament is forming at least one substantial star and could perhaps be thought of as a proto-stellar massive star forming core, however as we shall discuss later, there are also smaller cores within them, forming lower mass stars.

In the bottom two rows of \fig \ref{app} we show the central region of clumps Beta and Gamma. These clumps have been formed by several shock fronts intersecting to form a region of high density. Again, over time the extended structure collapses to form a more compact object. The density peaks, where the stars are forming, are carried along with this collapse. In effect we see the clump being formed at the same time as the stars are formed. As before, the clumps evolves from a diffuse filamentary distribution to a more centrally condensed distribution.

\begin{figure}
\begin{center}
\begin{tabular}{c}
\includegraphics[angle=0,width=2.2in]{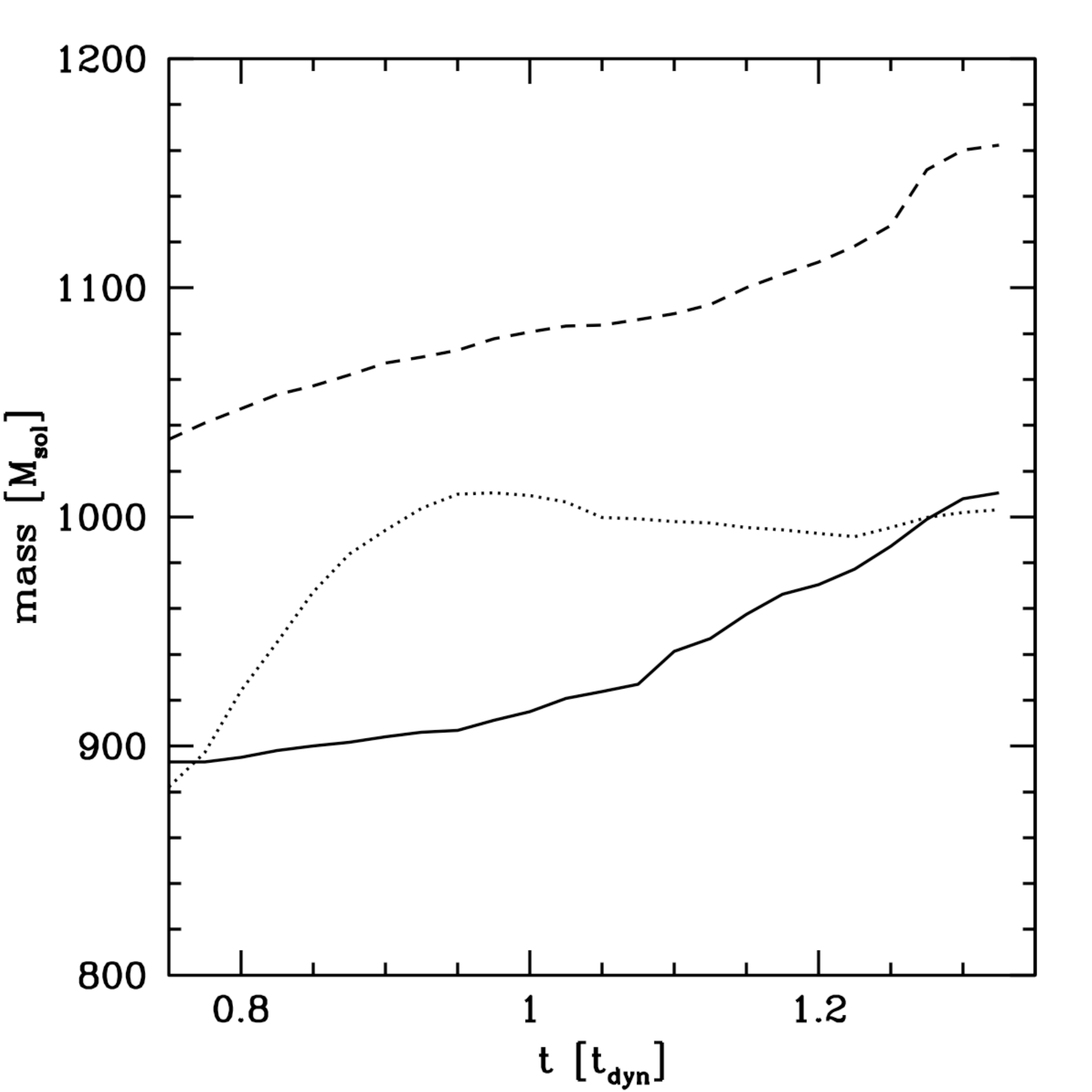}\\
\includegraphics[angle=0,width=2.2in]{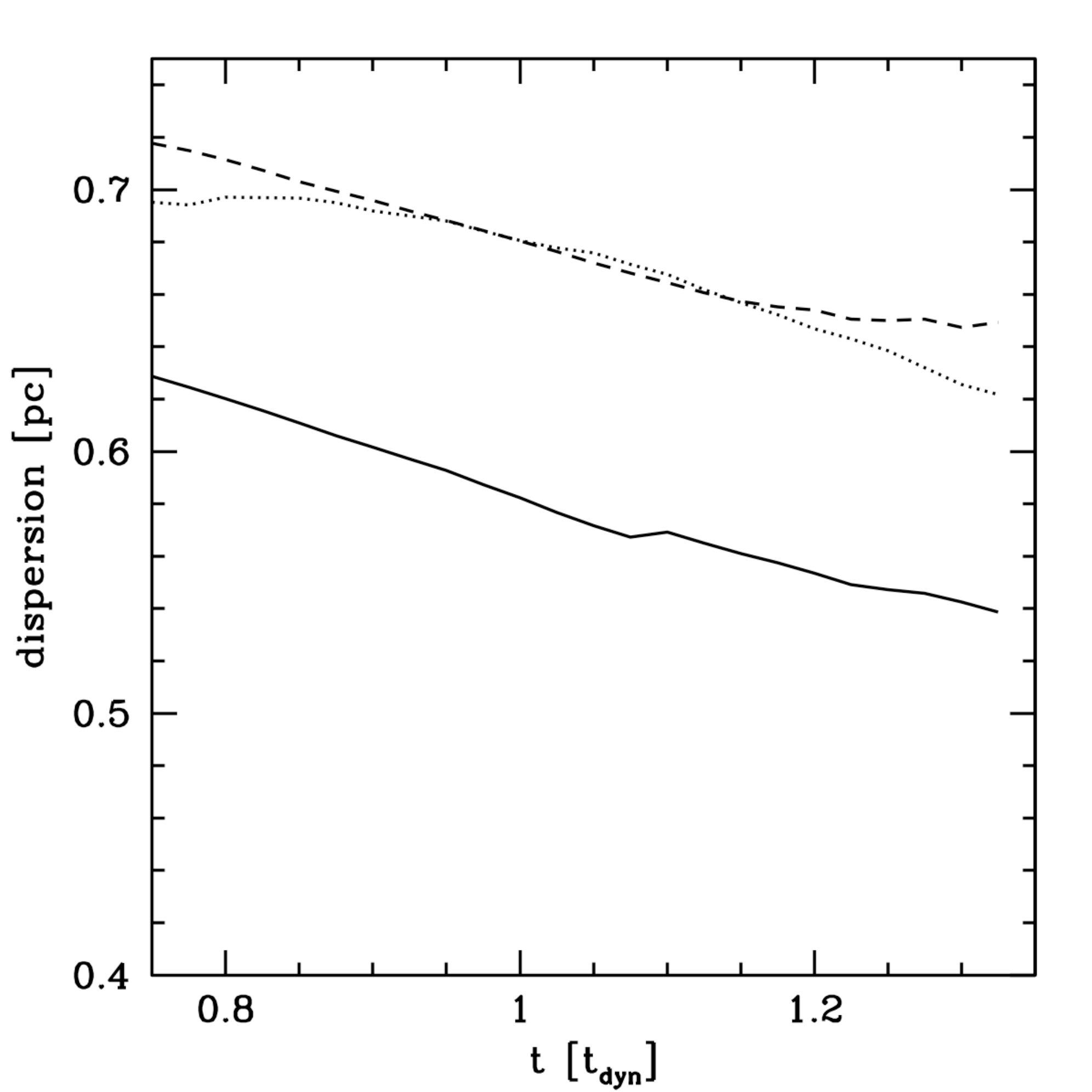}\\
\end{tabular}
\end{center}
\caption{The global properties of the mass within $1$ pc of the most massive sink for clumps Alpha \textit{(solid line)}, Beta \textit{(dotted line)} and Gamma \textit{(dashed line)} . \textit{Top}, the total mass and \textit{bottom}, the mass weighted dispersion of matter plotted against the simulation dynamical time (t$_{dyn}$ $\sim4.7\times10^{5}$ yrs). The clump mass increases with time and becomes more concentrated.}
\label{3Dprops}
\end{figure}

To describe more quantitatively the evolution of the clumps as a whole, we calculate two quantities using the full 3D data at each simulated time-step. Firstly, we record the absolute mass contained within a parsec radius of the most massive sink particle. Secondly, we calculate the dispersion of the surrounding mass using,
\begin{equation}\label{massdisp}
\centering
\sigma(r)_{3D}=\sqrt{ \frac{\sum m_{i}(r_{i}-\bar{r})^{2}}{\sum m_{i}} }
\end{equation}
where $m_{i}$ is the mass of the $i$'th SPH particle, and $r_{i}-\bar{r}$ is its distance from the central sink. Both gas masses and sink masses are included in this calculation. Figure \ref{3Dprops} shows the results.
The mass in the locality of the massive sinks generally increases with time. The exception to this was clump Beta where not all of the mass entering the region was bound and some escaped again. Moreover, the continuously decreasing dispersions show that the clump is collapsing and therefore continually channeling mass inwards. 
%Maybe comparison to the ONC here

\begin{figure*}
\begin{center}
\begin{tabular}{c c c}
\includegraphics[angle=0,width=2.2in]{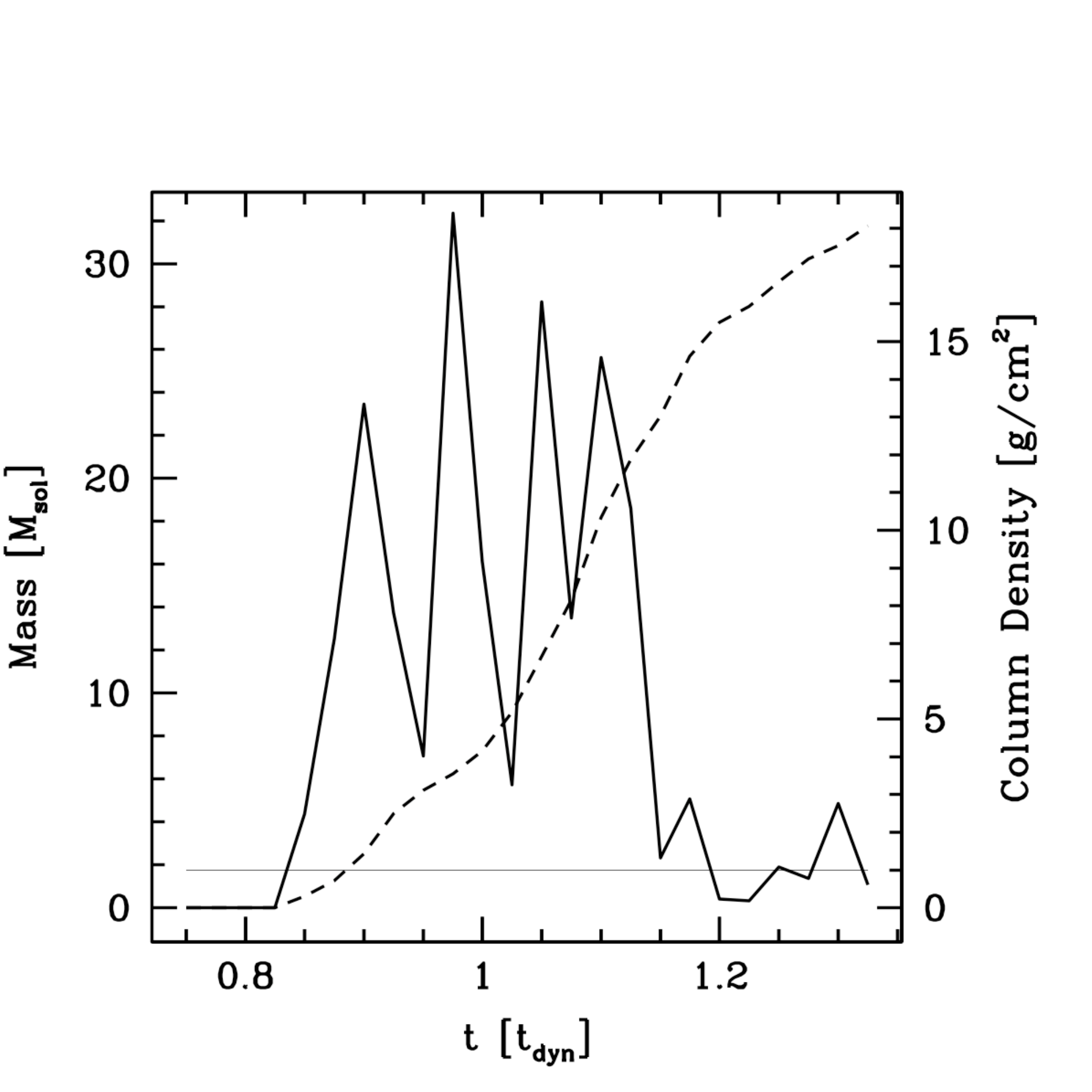}
\includegraphics[angle=0,width=2.2in]{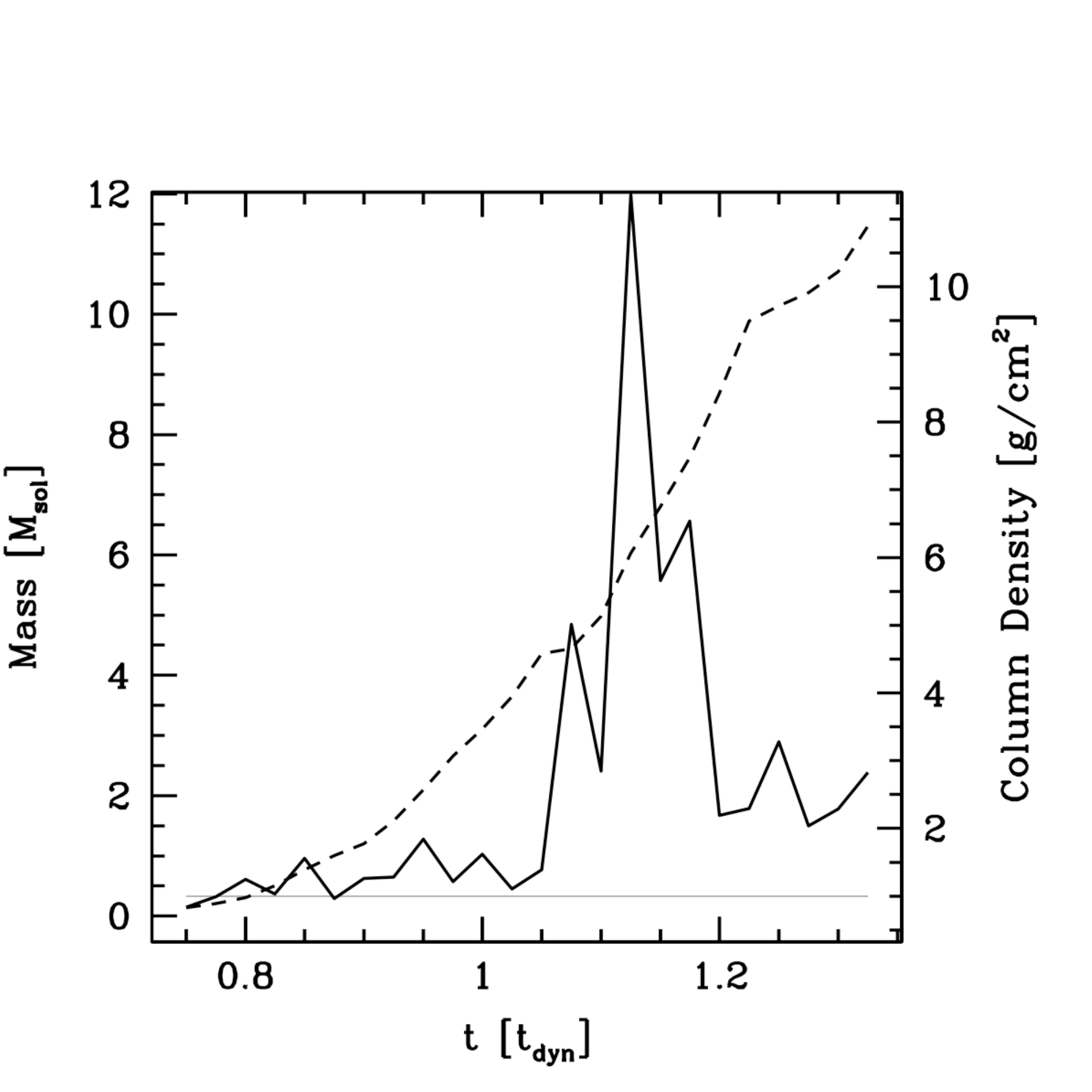}
\includegraphics[angle=0,width=2.2in]{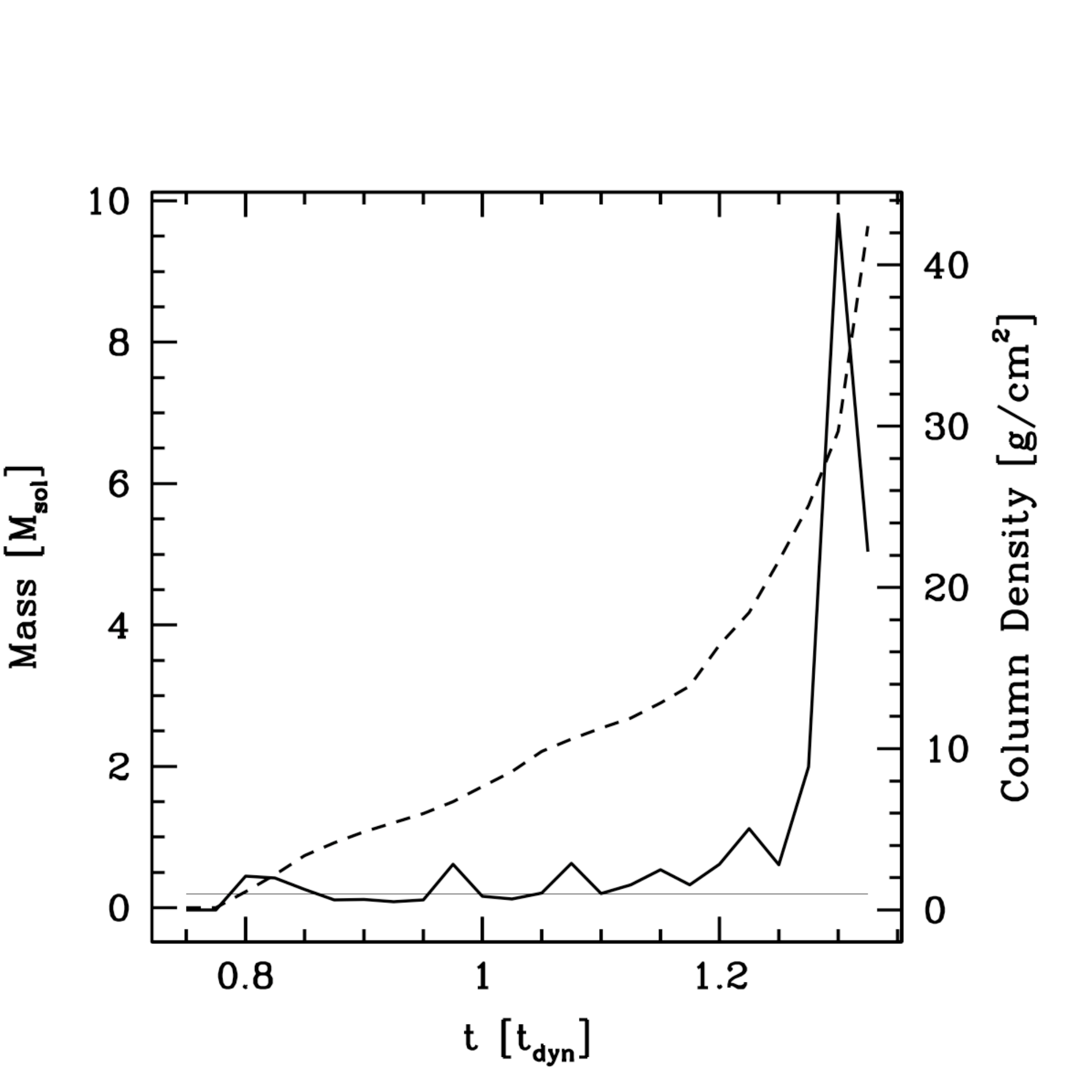}\\
\end{tabular}
\end{center}
\caption{The masses and column densities around the central sink in clumps Alpha \textit{left}, Beta \textit{middle} and Gamma \textit{right} plotted against the simulation dynamical time ( t$_{dyn}$ $\sim4.7\times10^{5}$ yrs). The dashed line shows the sink mass and the solid line the column density calculated in a $0.15$ pc box centered on the sink.}
\label{Coldens}
\end{figure*}

\citet{Krumholz08} found that a minimum gas column density of $1$ gcm$^{-2}$ was required for there to be sufficient feedback to avoid fragmentation and form a massive star. In Figure \ref{app} all regions coloured in yellow have column densities above $1$ gcm$^{-2}$, and this is where the majority of star formation is taking place. To illustrate this further, in Figure \ref{Coldens} the column density around the most massive sink in each clump is traced with time.

The column density around the central sink is calculated within a $0.15$ pc box, which is taken as a reasonably typical size for a massive proto-stellar core. The column densities are calculated from material within the 3D clump radius, as anything outwith this region cannot be affected by the forming star. However, as shown in Table \ref{Regprops} contamination from material along the line of sight could increase these values by up to a factor of two. On the same plot as the column density the growth of the central sink is shown. In all cases it increases to form a massive star. The column densities surrounding the sinks are consistently above the $1$ gcm$^{-2}$ limit in the calculated region. However, the average column density of the clump as a whole is an order of magnitude below this value. Although the column density surrounding the star was above the threshold proposed by Krumholz, we shall show in Section \ref{clump-core} that the massive stars were not formed from a single massive thermally supported fragment, but instead from a smaller core which accreted additional material channeled towards it by the potential of the forming stellar cluster. The high column densities, in this instance, seem mainly an indication of there being a large gas reservoir available for accretion.

\subsection{Observable Properties}

\begin{figure}
\begin{center}
\begin{tabular}{c}
\includegraphics[angle=0,width=2.5in]{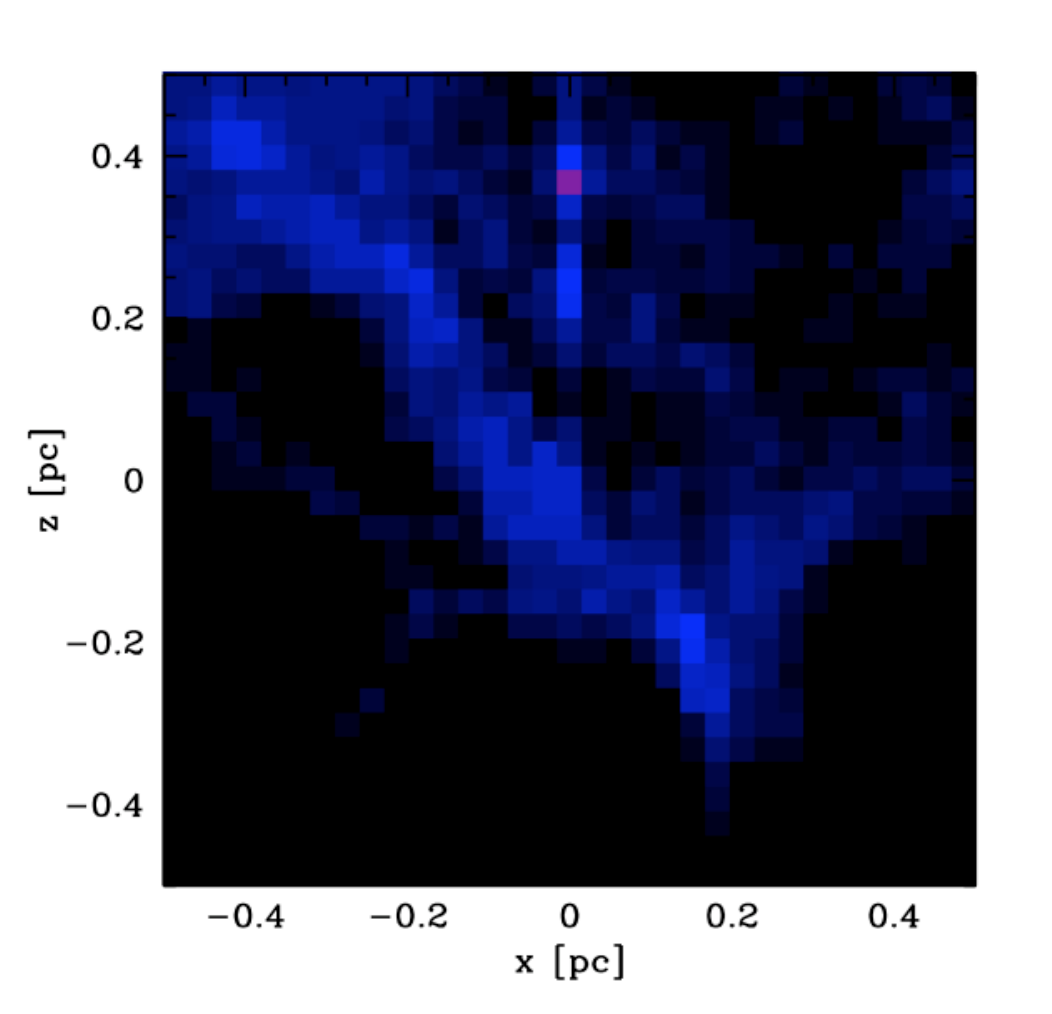}\\
\includegraphics[angle=0,width=2.5in]{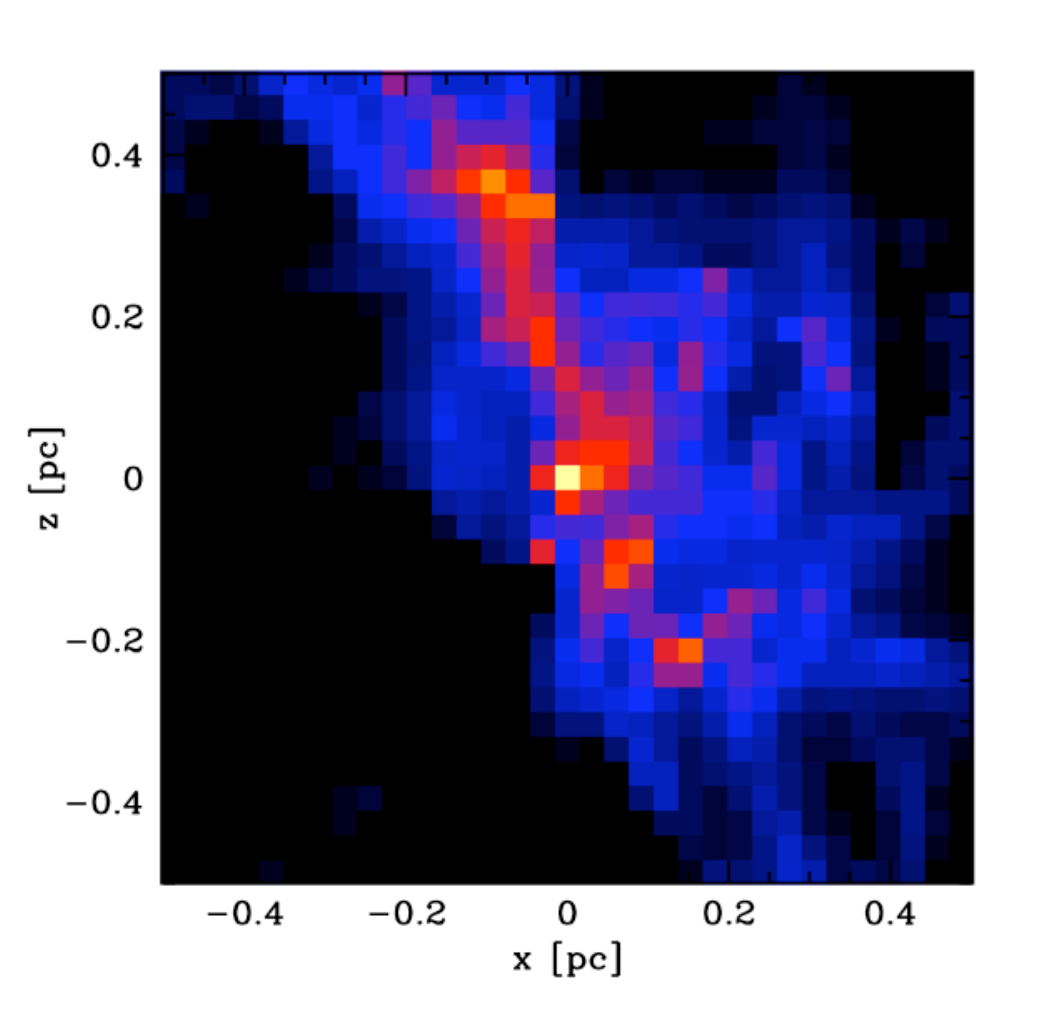}\\
\end{tabular}
\end{center}
\caption{The dust continuum emission at $230$ GHz from clump Alpha at \textit{top} $0.75$ t$_{dyn}$ ($3.53\times10^{5}$ yrs), and  \textit{bottom} $1.25$ t$_{dyn}$ ($5.9\times10^{5}$ yrs). The colour scale denotes emission from $0.5$ mJy (dark blue) to $500$ mJy (yellow). As the clump becomes more evolved the emission from its centre, where the massive stars are forming, increases.}
\label{CCemission}
\end{figure}

The above analysis uses the full 3D data-set. However, a better comparison to observations can be made by generating and analysing synthetic dust continuum emission images. To create an observers' version of \fig \ref{3Dprops} we interpolate our simulated data (including all material along the column) to a 2D grid of $66\times66$ grid cells, with a spatial resolution of $0.03$ pc, comparable to recent observations \citep[e.g.][]{Zhang09,Longmore09}. We then calculate the flux from each
grid cell using the relationship
\begin{equation}
F(\nu)=\sum_{i=1,n} {\frac{m_{i} g \kappa_{\nu} B_{\nu}(T_{i})}{d^{2}}}
\end{equation}
where $F(\nu)$ is the flux in Jy, $m_{i}$ is the mass of the SPH particle, $g$ is the dust to gas ration, $\kappa_{\nu}$ is the dust opacity, $B_{\nu}$ is the intrinsic emission of the SPH particle at temperature $T_{i}$ according to the Planck equation and $d$ is the distance at which the cloud is observed. We take the standard value of $0.01$ for the dust to gas ratio \citep{Kauffmann08} and a value of $0.1$ m$^{2}/$kg for the dust opacity \citep{Ossenkopf94}. The flux is calculated at a frequency of $230$ GHz ($1.3$ mm) and distance of $5$ kpc but these values do not affect the dispersion trends discussed below. We neglect emission directly from sink particles, but they still contribute to their surrounding gas particles emission due to the heating described in Section \ref{sec:sim}. Figure \ref{CCemission} shows the emission in mJy from Clump Alpha at early and late times. The emission increases with time, particularly in the centre of the clump where the massive star is forming.

\begin{figure}
\begin{center}
\begin{tabular}{c}
\includegraphics[angle=0,width=2.2in]{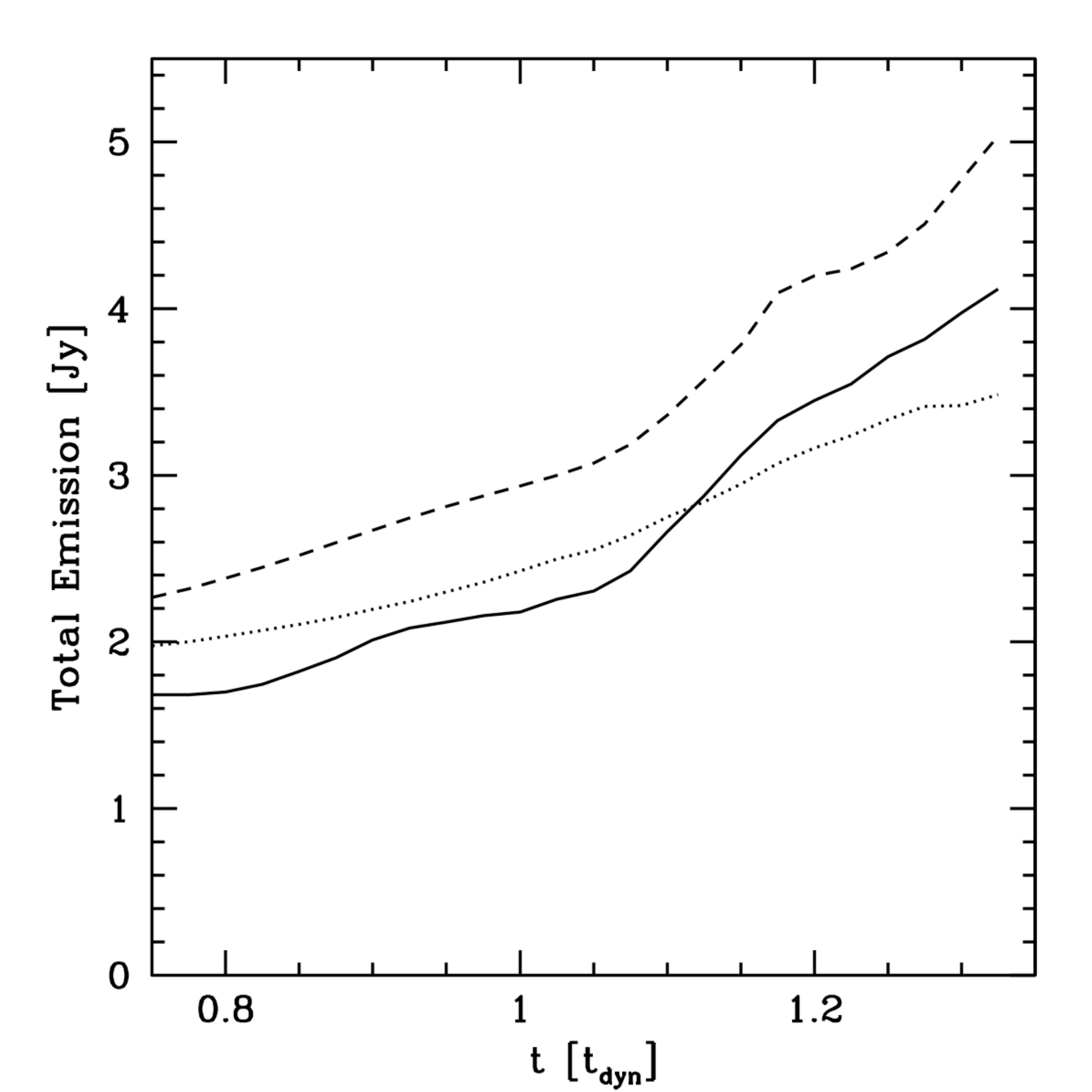}\\
\includegraphics[angle=0,width=2.2in]{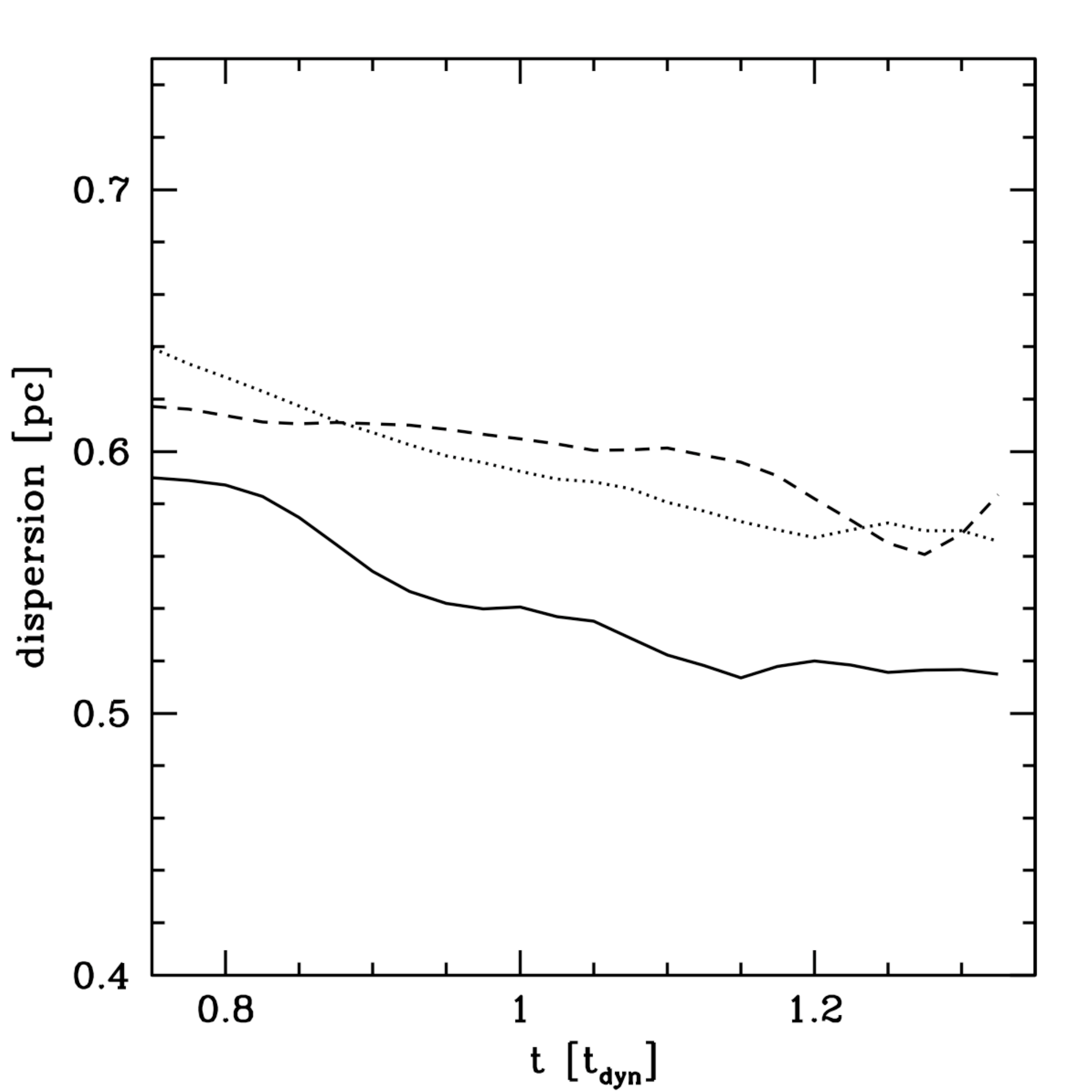}\\
\end{tabular}
\end{center}
\caption{The observable properties of the mass within $1$ pc of the most massive sink calculated from a 2D grid with a size resolution of $0.03$ pc, plotted against the simulation dynamical time (t$_{dyn}$ $\sim4.7\times10^{5}$ yrs). The clumps are denoted by the following lines; Alpha \textit{(solid line)}, Beta \textit{(dotted line)} and Gamma \textit{(dashed line)}. The panels show: \textit{top}, the total emission from the clump and \textit{bottom}, the emission weighted dispersion.}
\label{2Dprops}
\end{figure}

As before, a dispersion is calculated from the grid cells weighted by emission as shown in \eq \ref{emmdisp}.
\begin{equation}\label{emmdisp}
\centering
\sigma_{2D}(r)= \sqrt{\frac{\sum \xi_{i}(r_{i}-\bar{r})^{2}}{\sum \xi_{i}}}
\end{equation}
where $\xi_{i}$ is the emission from the grid cell $i$, and $r_{i}-\bar{r}$ is the distance from the grid cell where the largest sink is located. \fig \ref{2Dprops} shows the total emission and dispersion of the clumps with time.

The total emission from the clumps roughly doubles over the time considered here ($2.35\times10^{5}$ yrs). This is partially due to the increased mass of the clumps, but also due to increased emission from warmer dense gas where star formation is occurring. As in the three dimensional case a decreasing trend is seen in the dispersion. However, it is slightly less marked in emission due to the decreased resolution and the fact that only collapse along one plane is visible.

\subsection{Direct Comparison to Observations}

We now seek to directly compare the evolution of molecular gas structure predicted above to observations. Longmore et. al 2009 (L09) recently observed 6 massive star formation regions at 3 different evolutionary stages prior to UCHII region formation using the
Submillimeter Array at 230GHz to image the thermal dust continuum emission. They found the dust continuum emission to be weaker and more spatially extended at early stages and becoming more centrally concentrated with time. As the global properties of the L09 regions are similar to those of Alpha, Beta and Gamma, this dataset offers an excellent opportunity for comparison with the simulations.

\begin{figure*}
\begin{center}
\begin{tabular}{c c c}
\includegraphics[angle=0,width=2.5in]{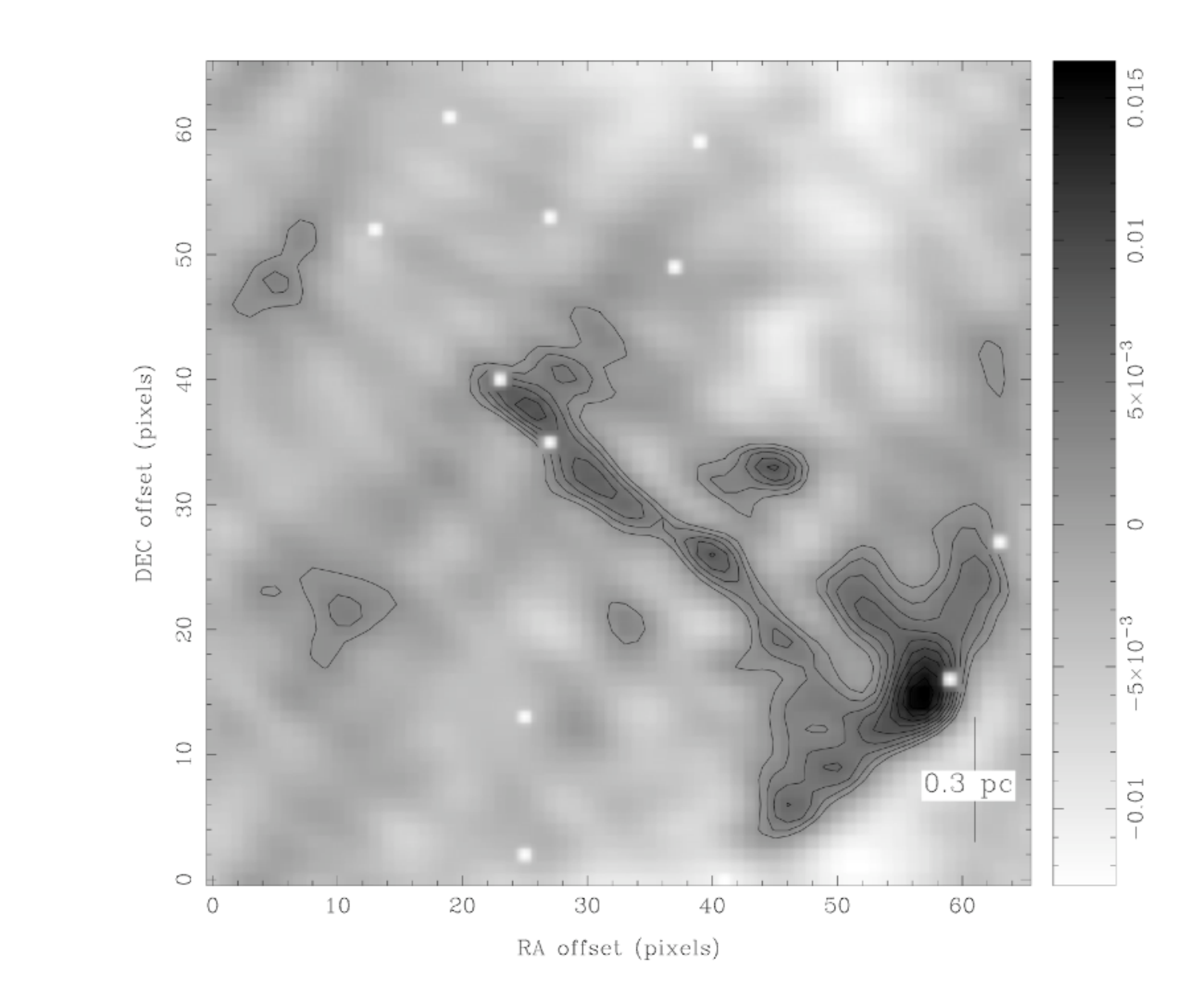}
\includegraphics[angle=0,width=2.5in]{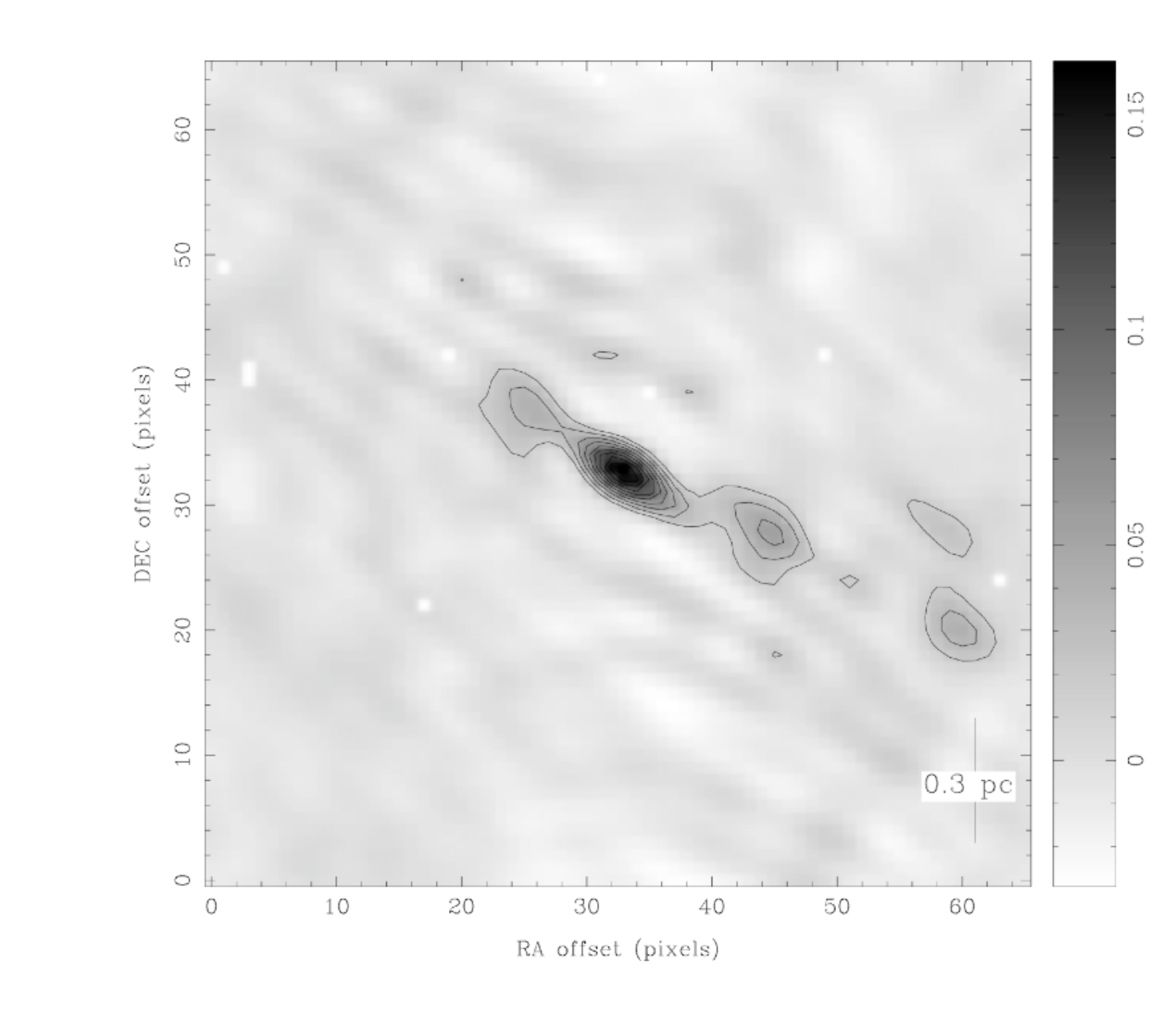}
\includegraphics[angle=0,width=2.5in]{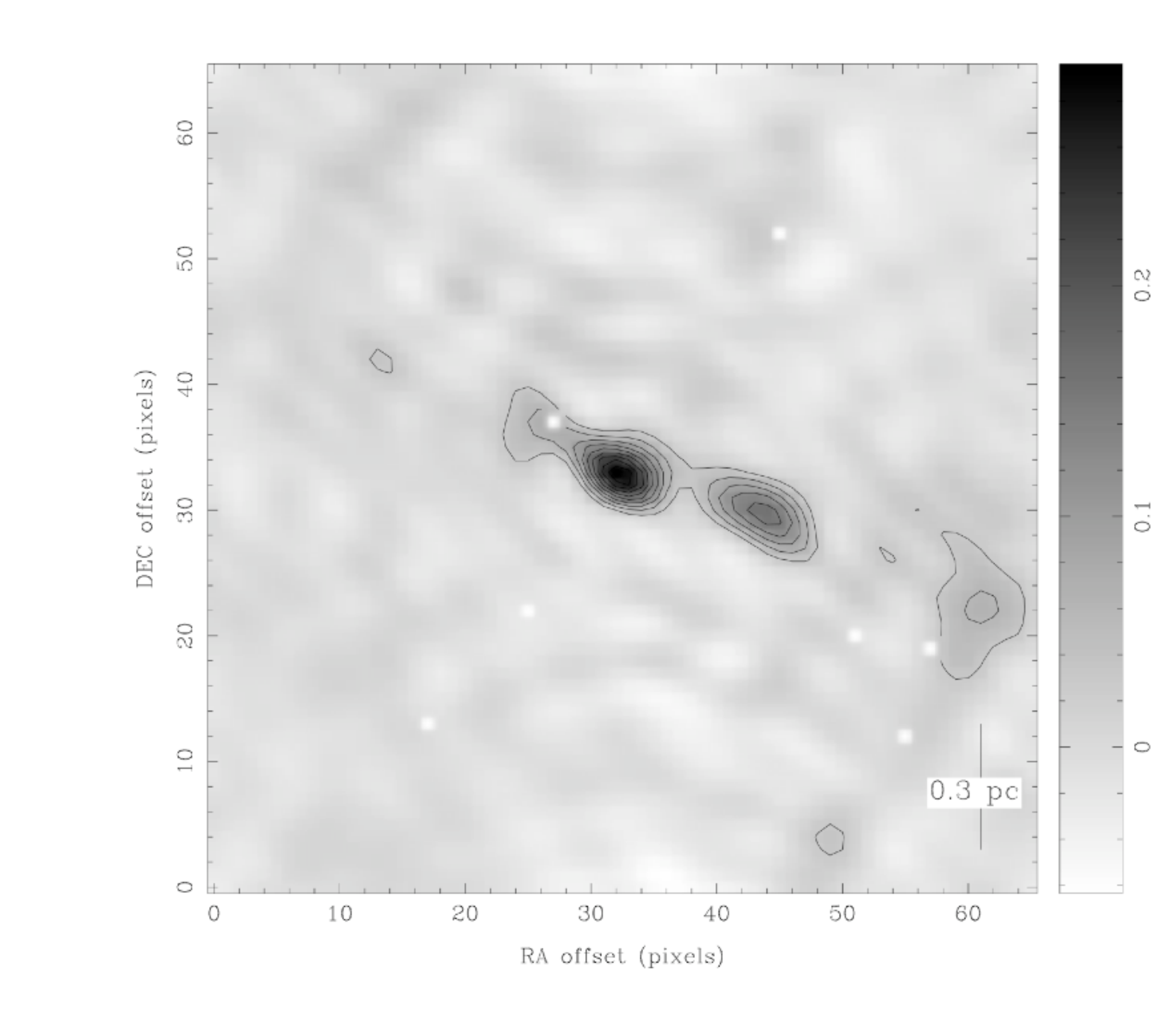}\\
\includegraphics[angle=0,width=2.5in]{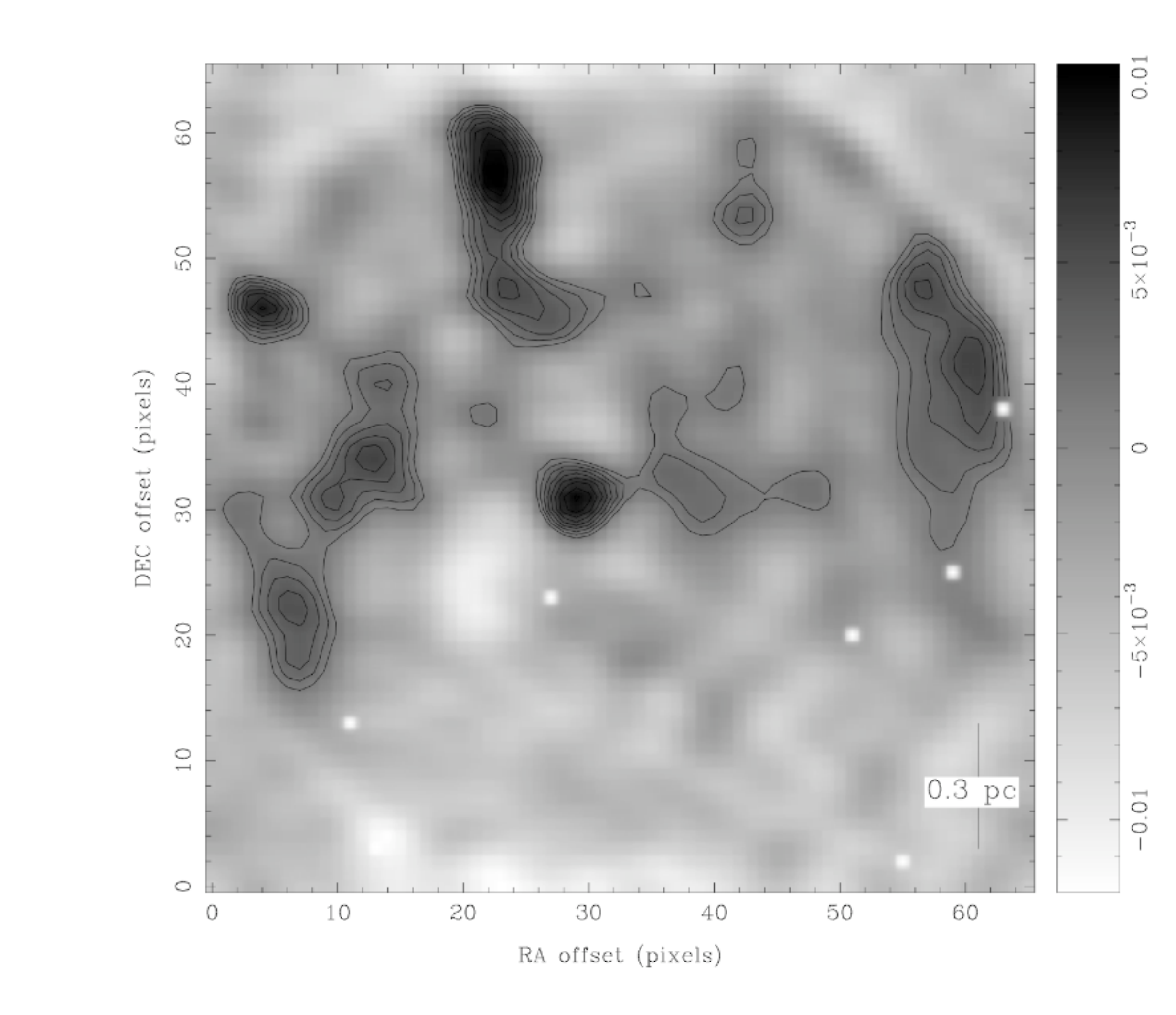}
\includegraphics[angle=0,width=2.5in]{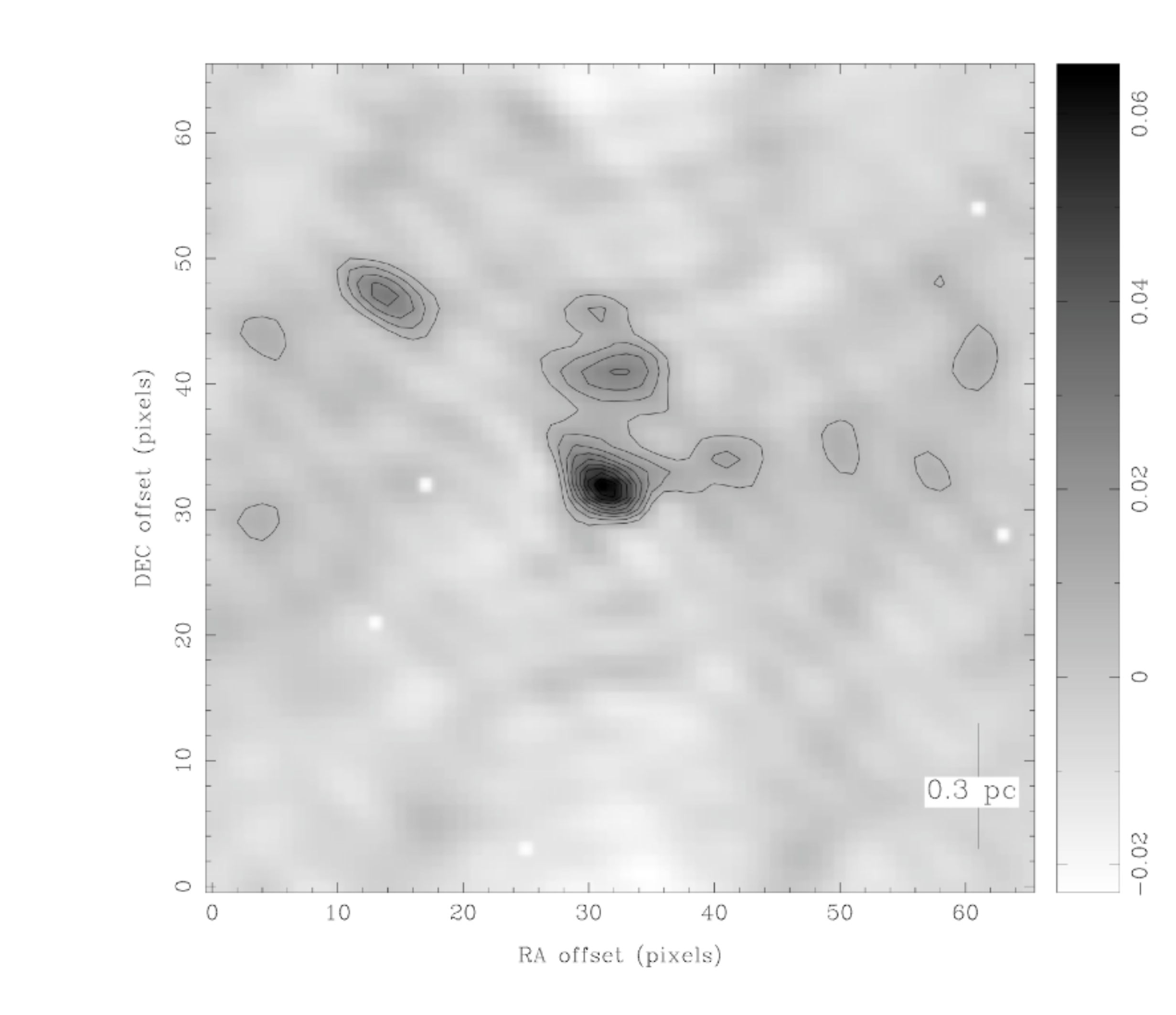}
\includegraphics[angle=0,width=2.5in]{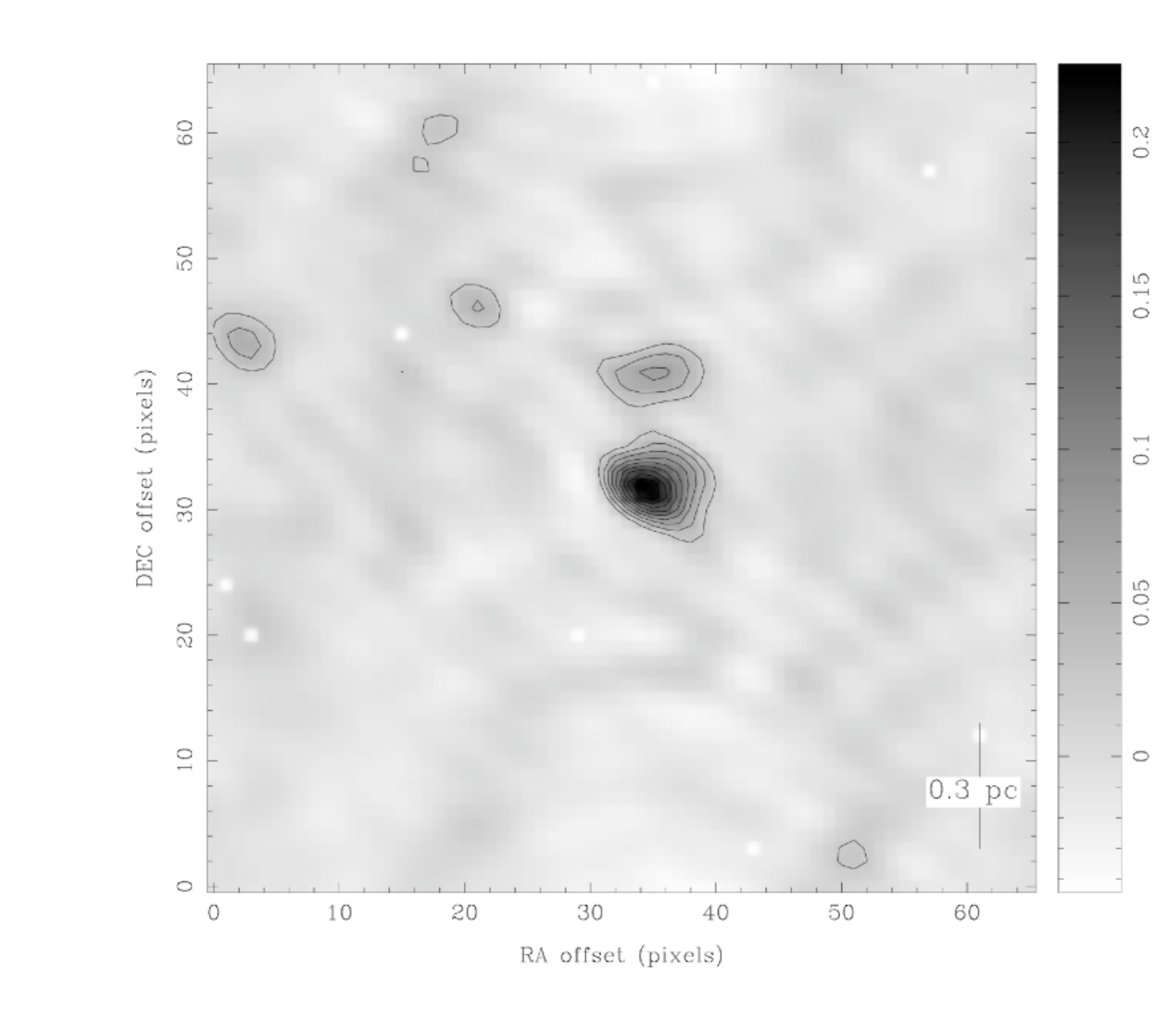}\\
\includegraphics[angle=0,width=2.5in]{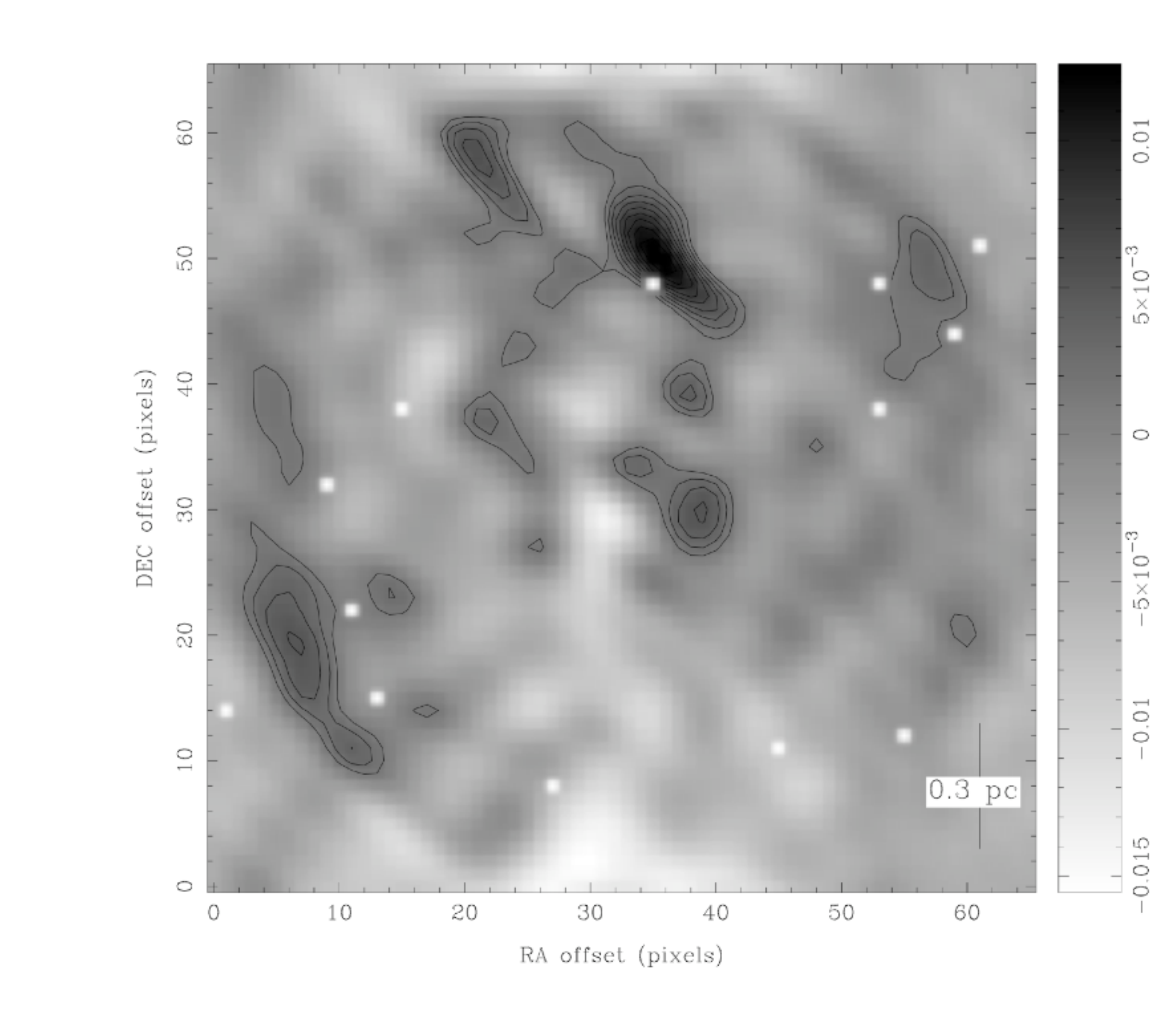}
\includegraphics[angle=0,width=2.5in]{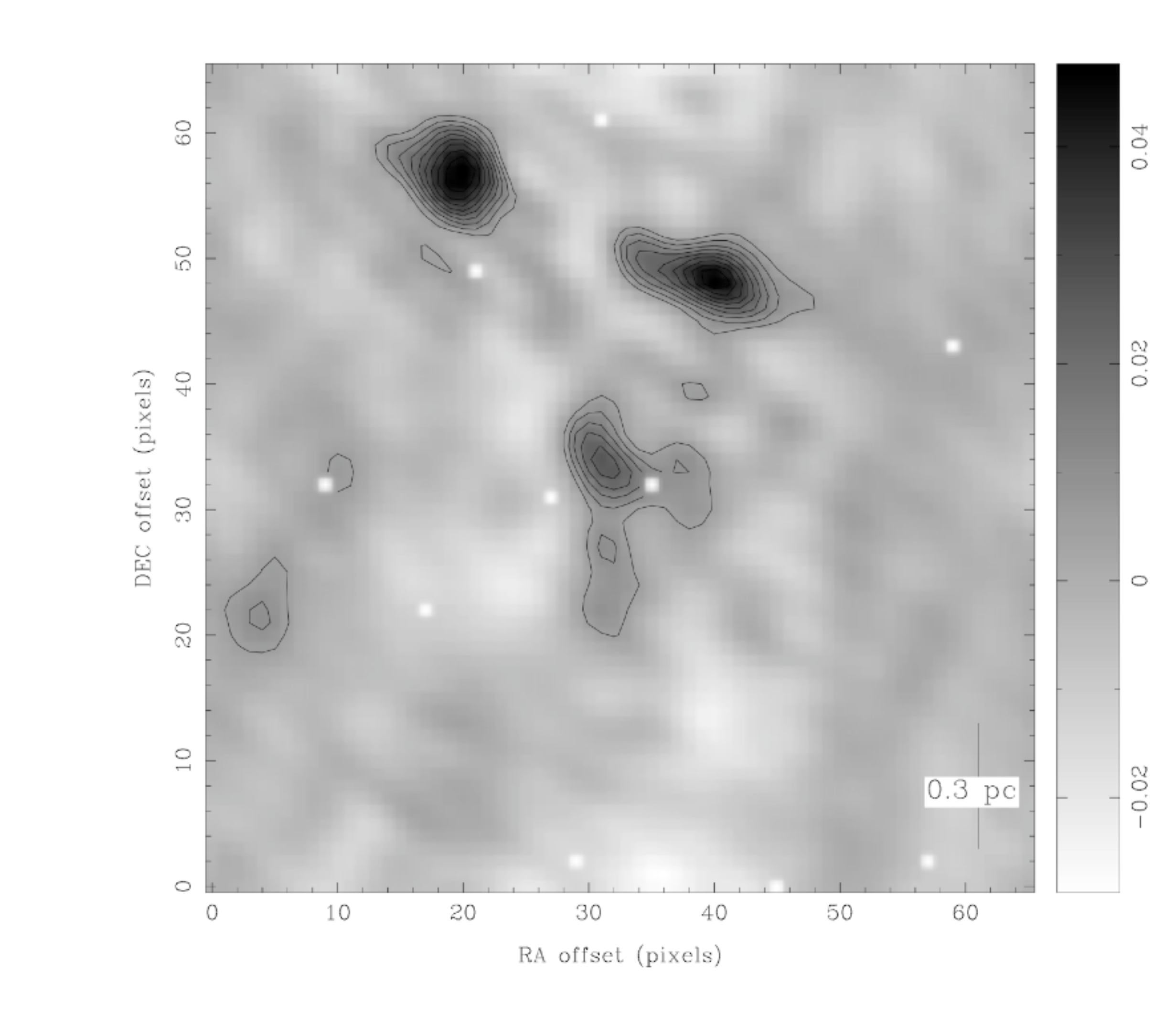}
\includegraphics[angle=0,width=2.5in]{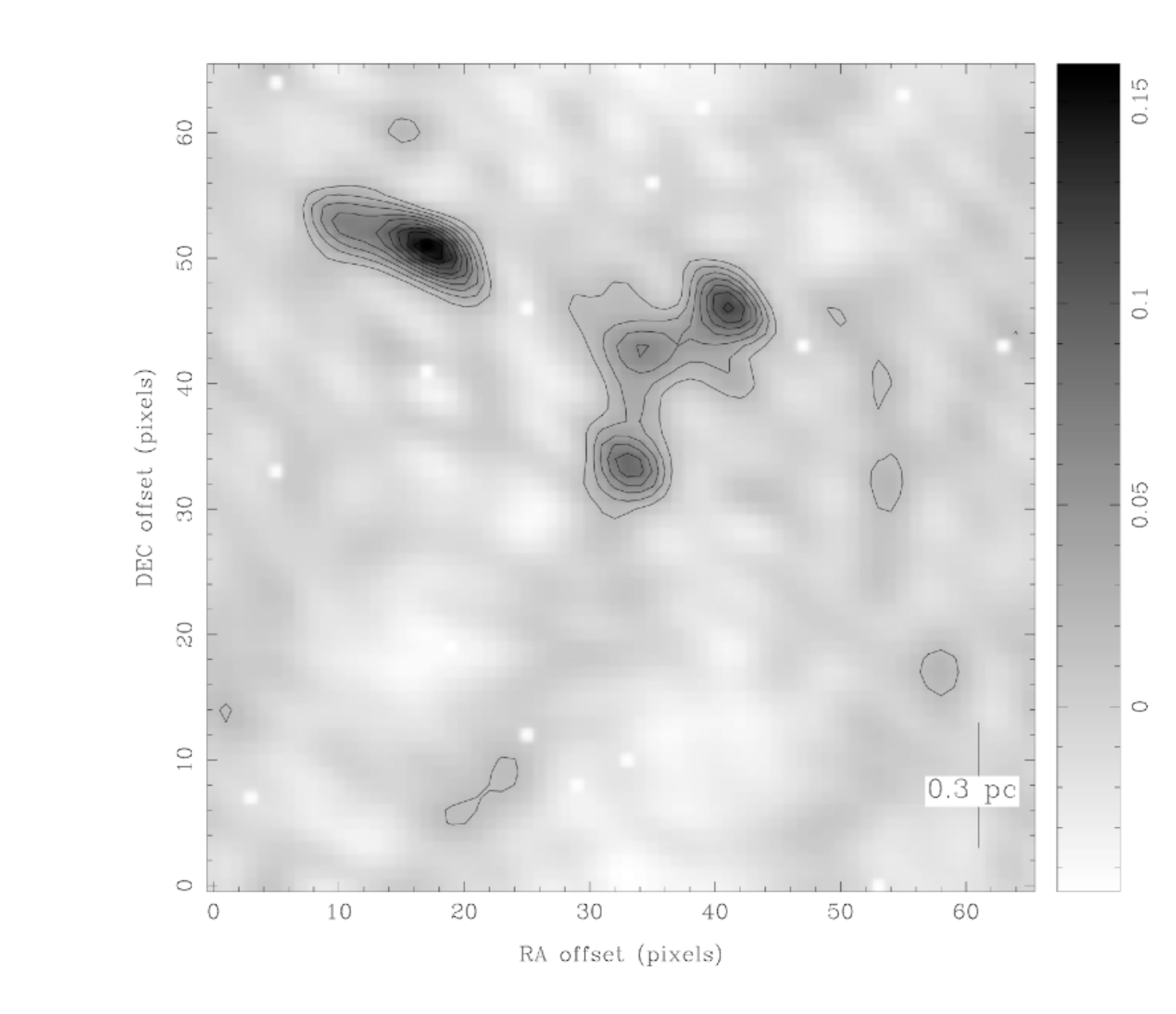}\\
\end{tabular}
\end{center}
\caption{Simulated interferometry observations of clumps Alpha \textit{top}, Beta \textit{middle} and Gamma \textit{bottom} at the same time intervals as shown in Figure \ref{app}.}
\label{inter}
\end{figure*}

The synthetic flux image was generated from the numerical simulation in the same way described above. To take account of the spatial filtering inherent in the interferometric observations, the simulated image was then sampled with the same uv-coverage as the L09 observations. The resulting synthetic 230GHz flux distributions towards Alpha, Beta and Gamma (at the same time steps as Figure \ref{app}) are shown in Figure \ref{inter}

The difference in source structure compared to Figure \ref{app} is striking. All of the largest scale extended emission has been
filtered-out and the images are instead dominated by the regions of highest density contrast. The much coarser resolution cannot
distinguish most of the fine detail, which is instead convolved into a smaller number of unresolved or partially-resolved sources.

The flux scale in each image, in units of Jy, is given by the colour bar. In all three regions the emission is a lot weaker at earlier times as the gas is more dispersed and cooler. Significantly more structure is seen at these earlier times down to the L09 sensitivity limit ($\sim$1 mJy) However, sensitivity is clearly an issue here -- more shallow observations would miss the weaker sources.

The contours in Figure \ref{inter} show the flux levels in the images in 10\% steps of the peak value. In all three regions, at early times the many sources in the field have similar flux densities. As the clump collapses its density decreases in its outer regions and increases at its centre. Moreover, the collapse feeds massive star formation at the centre of the clump, which heats the surrounding gas. This leads to the emission becoming dominated by one or two sources which are significantly hotter/denser than the surrounding cores. In real observations this may lead to dynamic range problems -- typical submm and mm interferometer images are limited to dynamic ranges of few hundred at best.

In summary, the synthetic dust continuum images closely match the observations of Longmore et al. (in preparation) and the numerical simulations offer an explanation for the trends seen in the spatial/flux distribution of the data.

\section{Discussion}
\label{sec:interaction}

\subsection{Collapse and Accretion}

\begin{figure}
\begin{center}
\begin{tabular}{c}
\includegraphics[angle=0,width=2.2in]{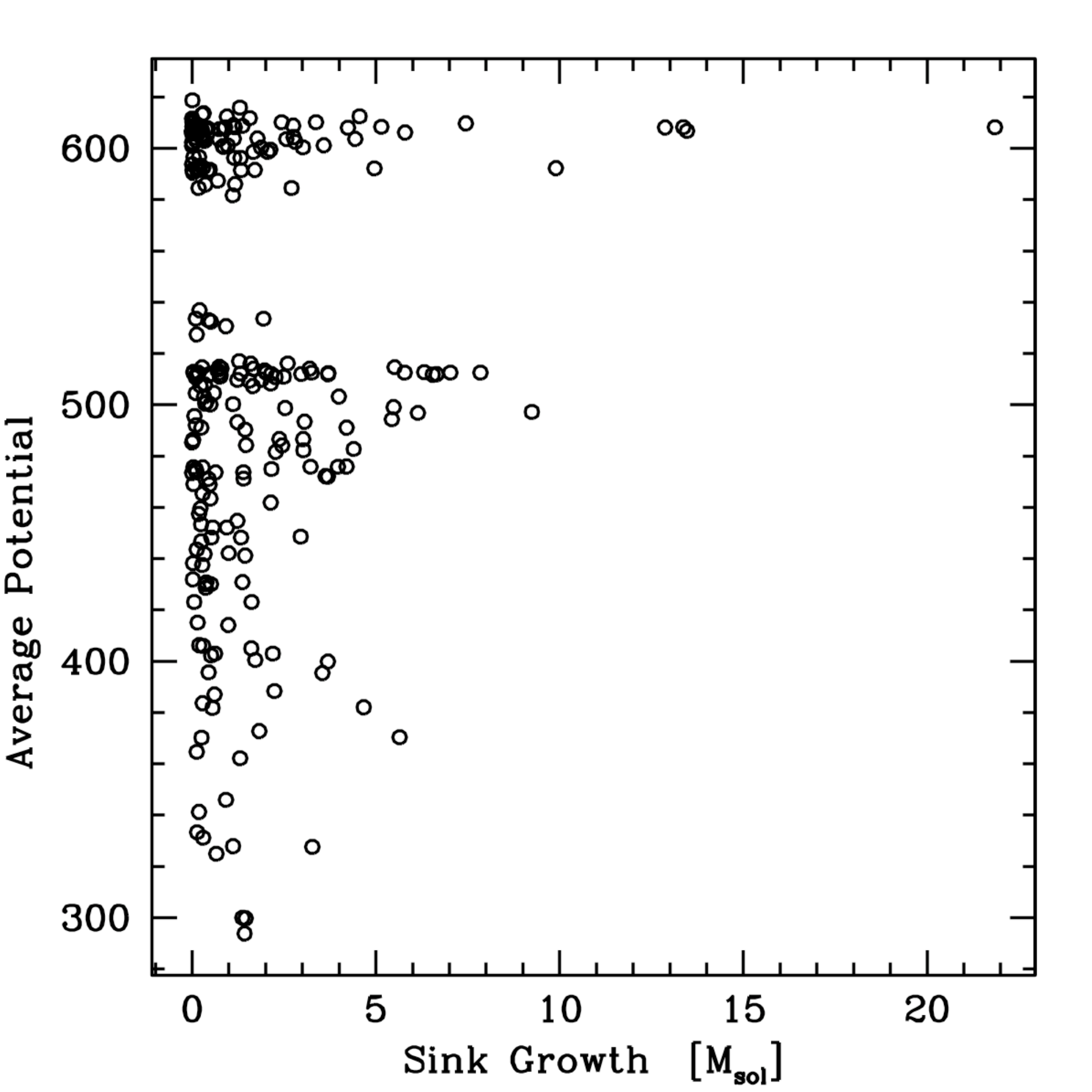}\\
\includegraphics[angle=0,width=2.2in]{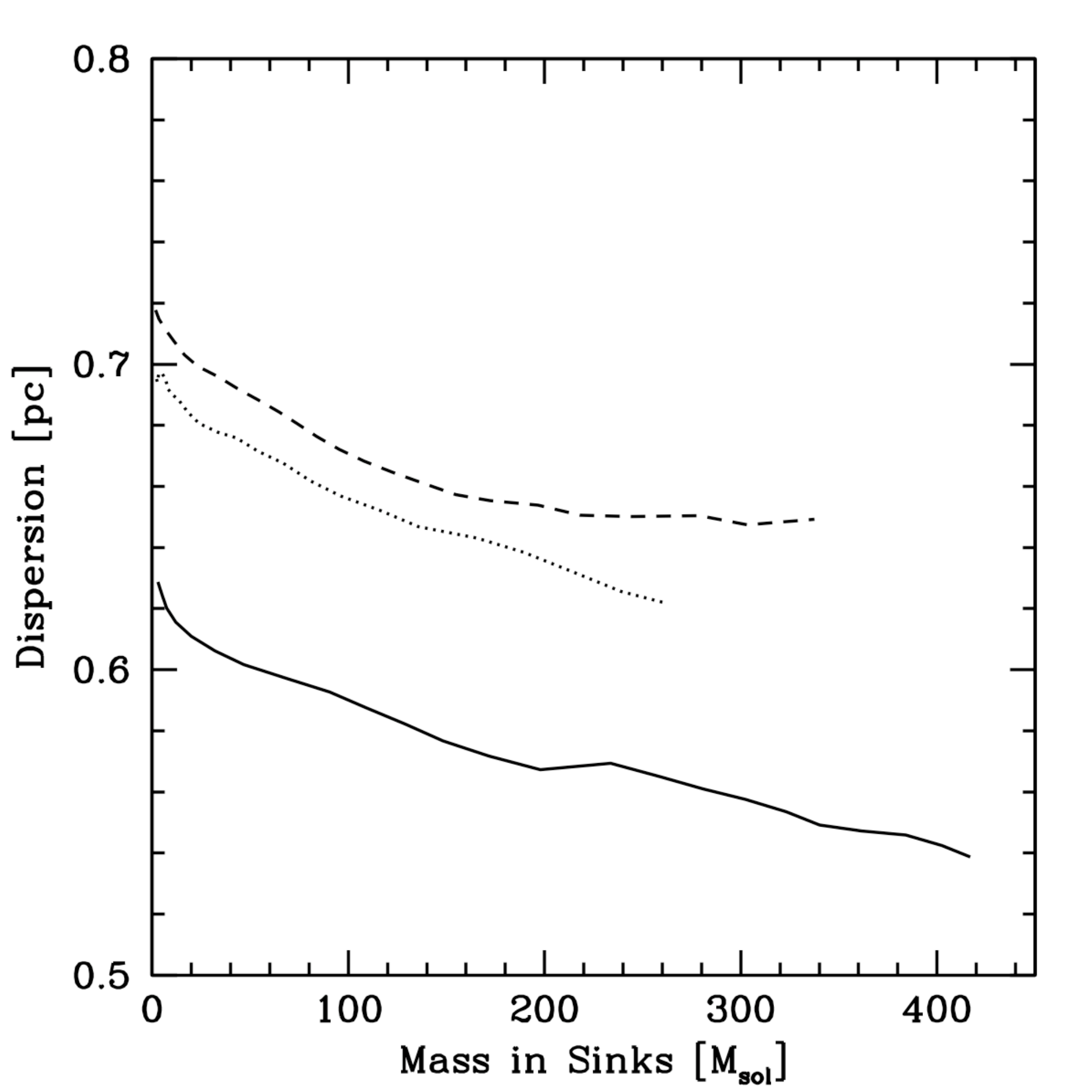}\\
\end{tabular}
\end{center}
\caption{The connection between clumps and sink masses during accretion. \textit{Top} the growth in sink mass over a period of $0.25$ t$_{dyn}$ plotted against the average potential in code units of the mass within a pc radius of the sink. The sinks in the deepest potential well grow the most significantly. \textit{Bottom} the mass in sinks within regions Alpha \textit{solid line}, Beta \textit{dotted line} and Gamma \textit{dashed line} plotted against their dispersion. The mass in sinks grows as the clump becomes more concentrated.}
\label{sinkgrowth}
\end{figure}

The evolution of the stellar cluster and the massive stars are intrinsically linked by the overall cluster potential and the accretion it induces. As the clump of gas evolves towards a stellar cluster it goes through several evolutionary stages. First, the clump becomes bound due to dissipation of turbulent energy, and the density enhancements within it begin to form bound cores. Secondly, as the clump becomes over-bound it undergoes global collapse which channels mass towards its centre, creating a large reservoir of gas. Thirdly, this gas will be accreted by the proto-stars with the largest accretion radii. \citet{Bonnell01} showed that for a collapsing system where the gas velocities and proto-stellar velocities are similar, the tidal radius is the most appropriate accretion radius. From analogy with the Roche Lobe formalism the tidal radius is,
\begin{equation}\label{rtide}
R_{acc}=C_{tidal}\left(\frac{M_{\ast}}{M_{enc}}\right)^{1/3} r_{\ast}
\end{equation}
where $M_{\ast}$ is the stellar mass, $M_{enc}$ is the mass enclosed within the cluster at the stars position $r_{\ast}$ and $C=0.5$ from the Roche Lobe approximation \citep{Eggleton83}. This gives a mass accretion rate of 
\begin{equation}\label{Mrate}
\dot{M_{\ast}} \approx \pi \rho v_{rel} R_{acc}^{2}
\end{equation}
where $\dot{M_{\ast}}$ is the accretion rate and $v_{rel}$ is the relative velocity between the proto-star and the gas. So the most massive star in a region is the most effective at gaining additional mass. There is also an additional boost to the accretion rate of the massive stars at the centre of the cluster from the enhancement of the local density by global in-fall.

The above analysis considers the initial evolution of the cluster when the gas and stellar velocities are well correlated. However, once the cluster becomes virialised and the gas and stellar velocities are no longer correlated, the massive stars will become even more efficient accretors \citep{Zinnecker82,Bonnell01}. 

Figure \ref{sinkgrowth} illustrates the connection between the clump collapse and accretion in our simulation. The top panel shows the growth in sink mass over a period of $0.25$ t$_{dyn}$ (t$_{dyn}$ $\sim4.7\times10^{5}$ yrs) plotted against the average eventual gravitational potential of the mass within a parsec radius of the most massive sink. Clump Alpha can be seen as a horizontal line of equi-potential along the top of the graph. The greatest growth in sink mass occurs in clump Alpha which has the deepest potential well. The greater potential shows that more mass has been concentrated in the centre, which acts to focus additional mass towards the massive sinks. This effect is not just due to limited numbers: there are $99$ sinks whose environment has an average potential above $550$ code units and $157$ below this value, but the three sinks which grow the most are all contained in the upper subset. This shows that the mass of the massive stars are linked to their environment. It is also worth noting that most of the stars in clump Alpha do remain as low-mass objects, and that these constitute the bulk of the stars within the stellar cluster formed from this clump. 

The bottom panel of Figure \ref{sinkgrowth} shows the total mass converted to sink particles in the three clumps plotted against their dispersion. The total mass in sinks increases as the clumps become more concentrated. In other words, the clump contracts and changes its distribution as it forms a stellar cluster. This evolution happens simultaneously with the evolution of the massive stars and affects their accretion.

\subsection{Clump-Core Interaction}
\label{clump-core}

\begin{figure*}
\begin{center}
\includegraphics[angle=0,width=3in]{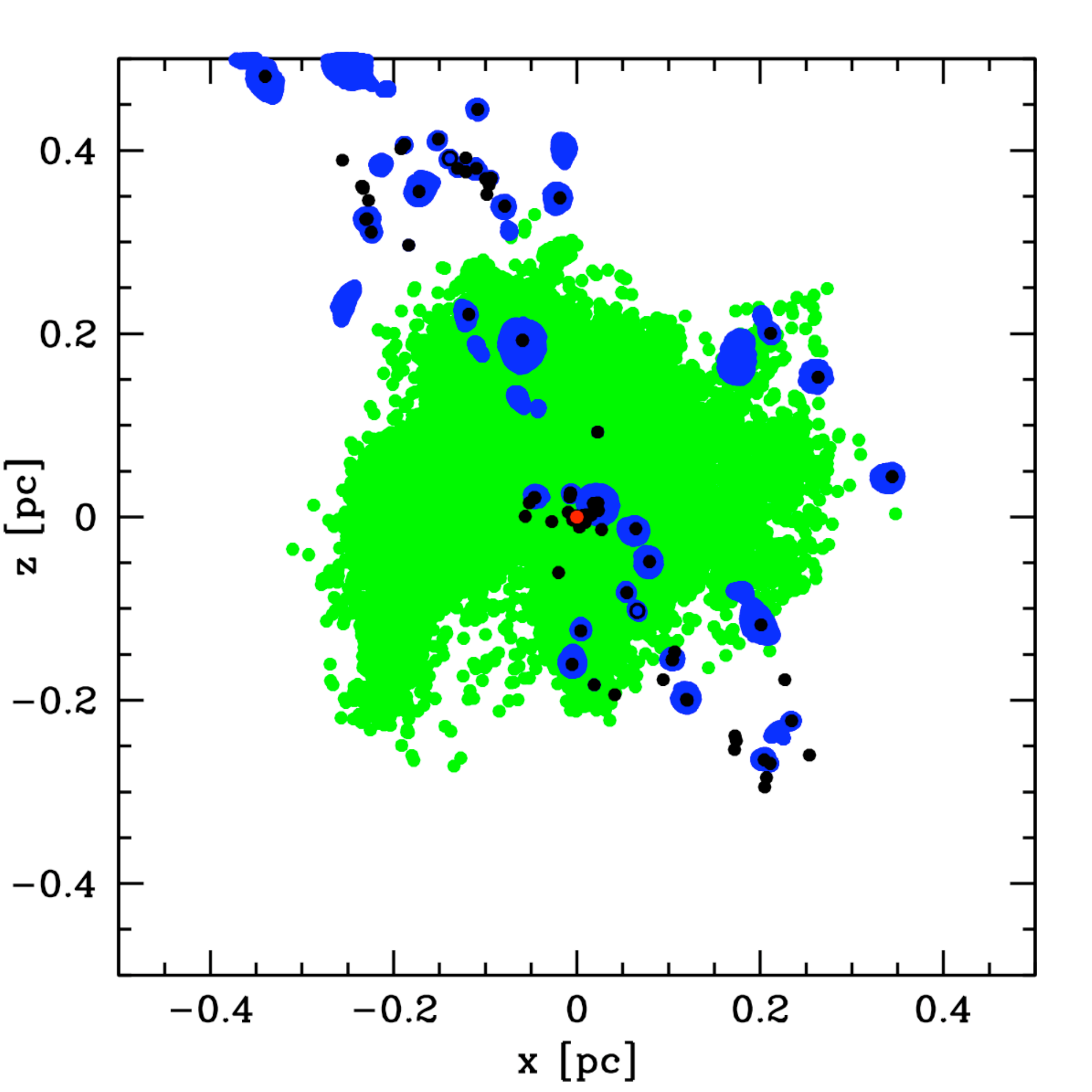}
\end{center}
\caption{The final fate of the mass within clump Alpha shown at $1$ t$_{dyn}$. The green dots show the positions of gas which will eventually be accreted by the massive sink (red dot). Black dots show the position of sinks and blue dots show the location of material in cores. The gas which will be accreted by the massive sinks is well distributed throughout the clumps, and generally cores within this region will not be disrupted by the massive sink. }
\label{massfate}
\end{figure*}

Next we consider how the clump and cores interact during global collapse. In \citet{Smith09} we demonstrated that the mass function of bound cores was indeed similar to the IMF, but the mapping between individual cores and their final stellar mass was poor. We concluded that this was mainly due to environmental factors. To investigate this further in Figure \ref{massfate}, we show the fate of the mass within clump Alpha at $1 t_{dyn}$. We use clump Alpha for illustration here as it has the simplest structure and contains the most massive sinks. The figure is colour coded to indicate the eventual fate of all the SPH mass particles at the end of the simulation. Green material will be accreted by the central massive sink (red dot). Black dots show the position of other sinks and blue regions show the location of material in gas cores. Our cores are `p-cores', which are identified using their gravitational potential wells as described in \citet{Smith09}. This method is less subjective than a typical CLUMPFIND, as it yields objects whose boundaries are not dependent on the contour numbers or density thresholds, and it ensures that there is a connection between the cores and the stars formed from them. In figure \ref{massfate} the blue cores will contain a black dot denoting a sink if they are proto-stellar, a hollow black dot if they are pre-stellar, and none if they are unbound.
% Maybe how many of each

\begin{figure}
\begin{center}
\begin{tabular}{c}
\includegraphics[angle=0,width=2.2in]{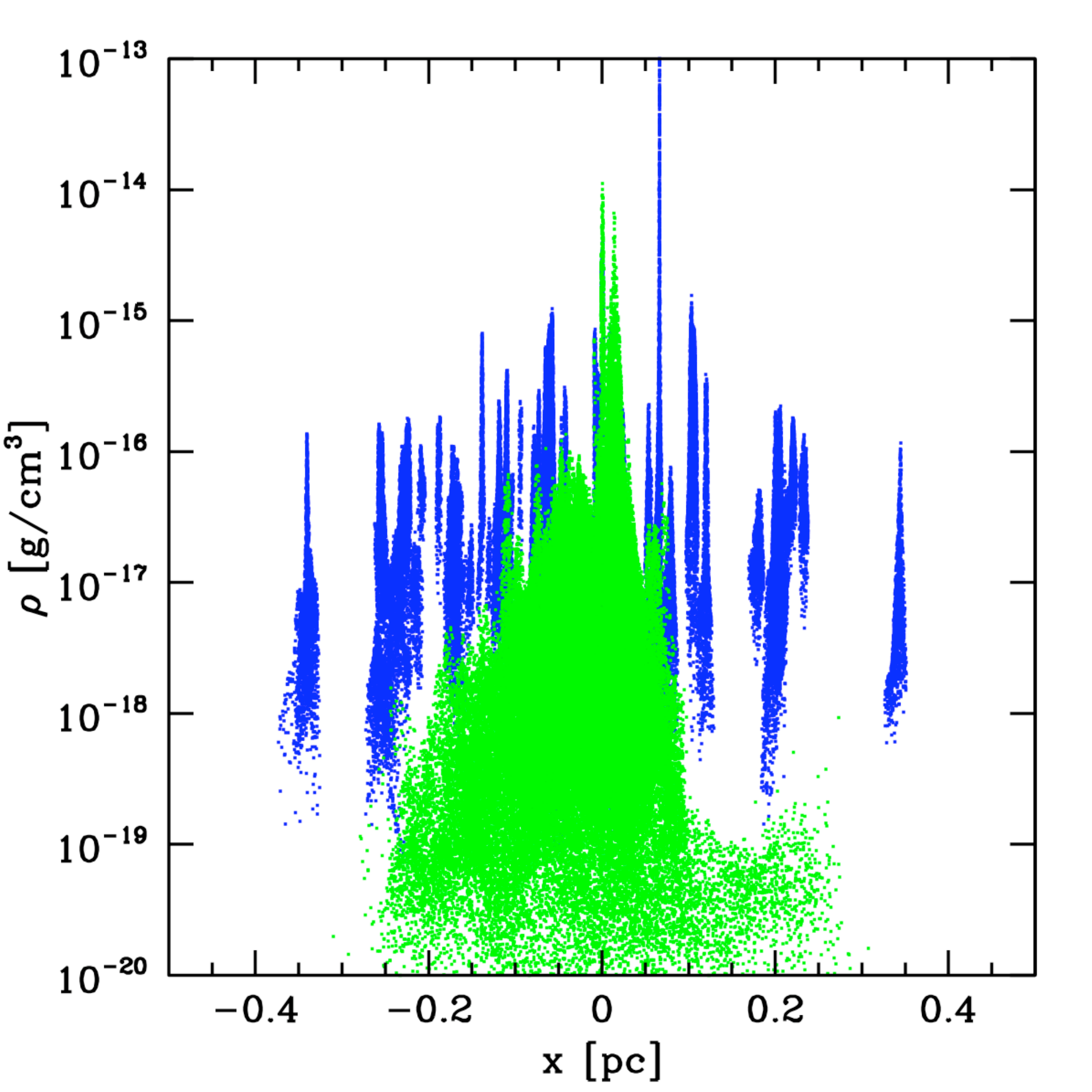}\\
\includegraphics[angle=0,width=2.3in]{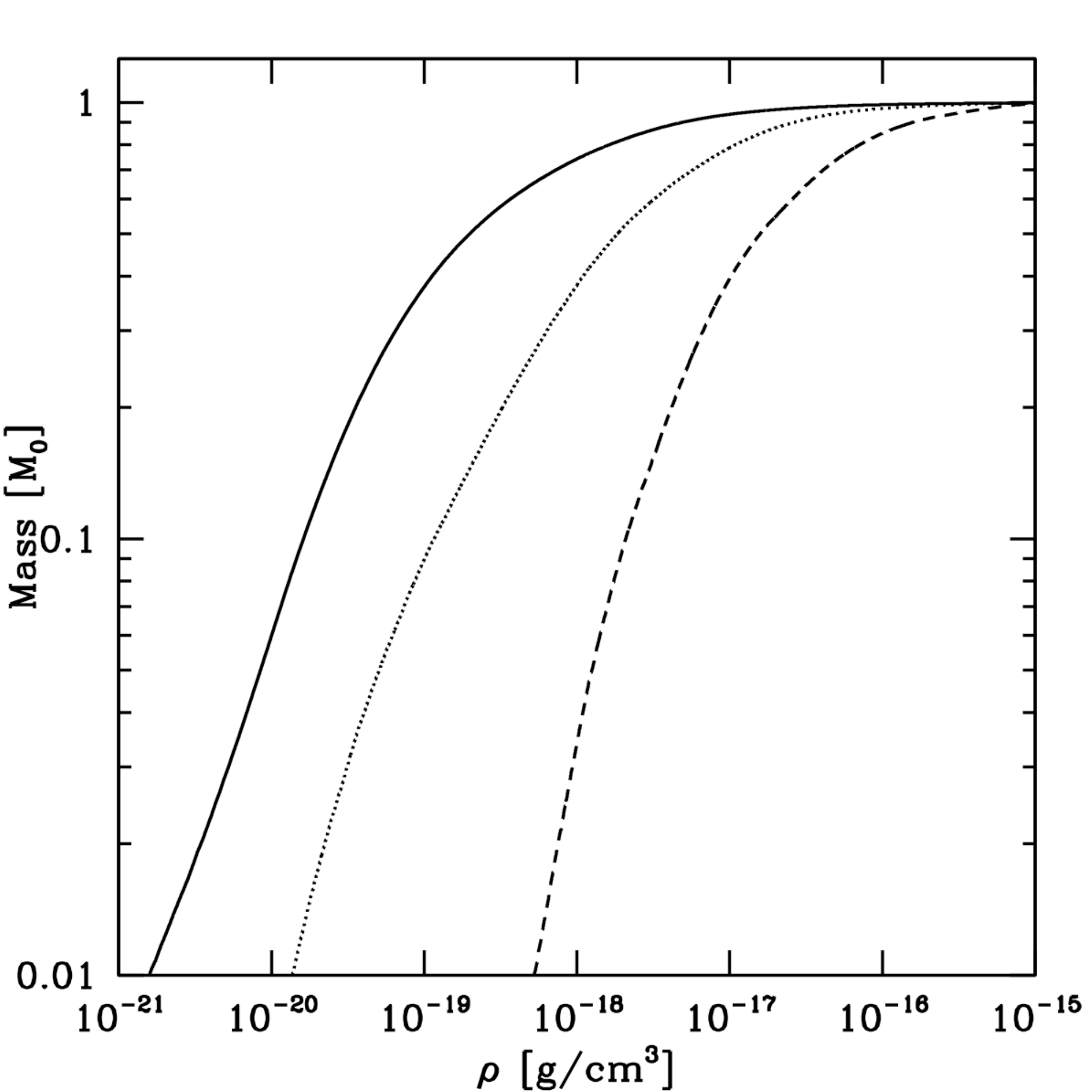}
\end{tabular}
\end{center}
\caption{\textit{Top} The density distribution of the material which will be accreted by the central sink in clump A (green), compared to that of the material in cores at $t_{dyn}=1$ (blue). \textit{Bottom} The cumulative density distribution of the entire clump (solid), material which will be accreted by the central sink (long dashed) and material in cores (short dashed). The masses have been normalised to magnitude one, to ease comparison. The material identified in cores contains mainly high density gas, whereas the material which will be accreted by the central sink contains significant amounts of low density gas.}
\label{massfatedens}
\end{figure}

The gas which will be accreted by the massive sink is well distributed throughout the clump, and it comes from a larger area than the typical size of a p-core. The p-cores sit within the volume from which mass will be accreted by the central massive sink, but are largely unaffected by this. Only the proto-stellar core directly to the left of the massive central sink gets disrupted due to heating. Additionally, an unbound core above the massive sinks is also destroyed and then accreted by the central sink before it can become bound. To illustrate why most of the low mass proto-stellar cores are unaffected by accretion from the central sink, we plot their respective densities in Figure \ref{massfatedens}. 

The top panel of Figure \ref{massfatedens} shows at t$_{dyn}=1$, the densities of sph particles which will later be accreted by the central sink in green, and the densities of the particles in cores in blue. The accreted material extends to low densities, whereas the gas in cores is confined to higher densities. As the free fall time of the gas is proportional to density as $t_{ff}\propto \rho^{-1/2}$, the cores have short dynamical times and can collapse before they can be accreted. The lower density gas between the cores has a longer free fall time and therefore can be accreted by the central sink. This can be seen in the way that the density of accreted gas decreases with distance from the massive sink, as it needs to have a longer free fall time in order to successfully reach the sink. 

In the bottom panel of Figure \ref{massfatedens} we show the gas density plotted against cumulative mass for clump Alpha as a whole, the accreted material, and the cores. The material in cores has characteristic densities, $\rho_{c} \sim 10^{-17} $ gcm$^{-3}$, the accreted material has $\rho_{c} \sim 10^{-18} $ gcm$^{-3}$, and the clump as a whole has $\rho_{c} \sim 10^{-19} $ gcm$^{-3}$. Once again, this shows that the accreted material has a wider initial density distribution range than the cores. It also shows that if the respective distributions were observed above a density threshold, the mass available for accretion would be underestimated. For instance, over $20\%$ of the mass accreted by the central sink has a density below the minimum seen in cores.

A major difference between this and other models of high mass star formation is that within a typical massive star forming clump, there are smaller proto-stellar cores forming low mass stars close to where the high mass star is forming ($r<0.15$ pc), and within the region which it accretes from. This would be most apparent at early times in the evolution of the clump when it is still diffuse, before the cores have become concentrated at the centre, and the emission becomes dominated by the central source.

\subsection{Massive Star Progenitors}

When the original progenitor of the massive sink was searched for using the p-core routine, it was found to be a bound core containing $0.67$ \msun. This is very close to the mean bound p-core mass of $0.7$ \msun, and is an order of magnitude lower than the final mass of $29.2$ \msun the sink achieved by the end of the simulation. Similarly, for the other clumps studied here, their most massive sinks are found to originate from intermediate-mass pre-stellar cores which become massive proto-stars via accretion. Therefore, the mass which forms the massive star comes mainly from the larger clump, rather than from a well defined massive pre-stellar core. In this scenario initially there is a low-to-intermediate pre-stellar core at the centre of the potential which grows to a massive condensation (YSO) as the clump collapses. It does not matter if smaller fragments form around this object, as the evolving clump continues to channel mass towards it.

\section{Conclusions}
\label{sec:conc}
We have carried out SPH simulations of a giant molecular cloud with a simplified radiative feedback model. From this simulation we identified three gas clumps of radius 1 pc which are the progenitors of stellar clusters. We find that the formation of a stellar cluster occurs simultaneously with massive star formation. The evolution of the two are therefore intrinsically linked. This leads to the following predictions.
\begin{enumerate}
\item Massive clump structure is originally diffuse and filamentary, but evolves into a more concentrated structure by means of gravitational contraction.
\item The models presented here are in good agreement to the interferometry observations of Longmore et al. (in preparation). Simulated interferometry images show more structure at early times, and less at later times when the emission is dominated by hot, dense central sources.
\item Both the most massive stars and the most massive stellar cluster are formed within the most bound clump. This is despite it being the least massive of the three clumps studied.
\item The gravitational potential of the gas clumps causes global collapse, which continuously channels mass from large radii towards the centre of the cluster, where it is accreted by the progenitors of the massive stars.
\item The original pre-stellar core of the most massive sink formed was only of intermediate mass. Most of the mass which goes into the sink originally came from the less dense clump gas between the surrounding low mass cores.
\end{enumerate}

\section{Acknowledgements}
We would especially like to thank Thomas Robitaille for the use of his models and his help in their implementation here. We would also like to thank Enrique Vazquez-Semadeni, Mark Krumholz, Paul Clark and Robert Smith for useful discussions.

\bibliography{Bibliography}

\label{lastpage}

\end{document}